\documentclass[12pt, draftclsnofoot, onecolumn]{IEEEtran}
\IEEEoverridecommandlockouts
\usepackage[T1]{fontenc}
\usepackage{graphicx}
\usepackage{cite}
\usepackage{subcaption}
\usepackage{caption}
\usepackage{algpseudocode}
\usepackage{lipsum}
\usepackage{overpic}
\usepackage{amssymb}
\usepackage{amsmath}
\usepackage{amsthm}
\usepackage{array}
\usepackage{color}
\usepackage{url}

 \newcommand{\sgn}{\text{sgn}}

\usepackage{optidef}
\usepackage{times}
\usepackage{fancyhdr,graphicx,amsmath,amssymb}
\usepackage[ruled,linesnumbered,noend]{algorithm2e}

\IEEEoverridecommandlockouts

\usepackage{cite}
\usepackage{graphicx}
\usepackage{url}
\usepackage{relsize}
\usepackage{hhline}
\usepackage{nicefrac}
\usepackage{amssymb,bm,upgreek}
\usepackage[table]{xcolor}
\usepackage{multirow}
\usepackage{algpseudocode}
\usepackage{subcaption}
\usepackage{algpascal}
\usepackage{amsthm}

\usepackage{caption}
\usepackage{xcolor}

\newtheorem{remark}{Remark}

\usepackage{optidef}
\usepackage{overpic}
\theoremstyle{remark}

\begin{document}

\title{Optimized Precoding for MU-MIMO With Fronthaul Quantization}

\author{\IEEEauthorblockN{\normalsize
    Yasaman Khorsandmanesh, \textit{Student Member, IEEE},  Emil Björnson, \textit{Fellow, IEEE}, and \\ Joakim Jaldén, \textit{Senior Member, IEEE}}
    \thanks{{
Y. Khorsandmanesh and E. Björnson are with the Division of Communication Systems, KTH Royal Institute of Technology, Stockholm, Sweden (E-mails: \{yasamank, emilbjo\}@kth.se). J. Jaldén is with the Division of Information Science and Engineering, KTH Royal Institute of Technology, Stockholm, Sweden (E-mail: jalden@kth.se).}}
\thanks{This work was supported by the Knut and Alice Wallenberg Foundation.}
\thanks{This work was presented in part at the IEEE International Conference on Acoustics, Speech, and Signal Processing (ICASSP), Singapore, May 2022, which appears in this manuscript as reference \cite{yasaman2022}.}

}

\maketitle 

\begin{abstract}
One of the first widespread uses of multi-user multiple-input multiple-output (MU-MIMO) is in 5G networks, where each base station has an advanced antenna system (AAS) that is connected to the baseband unit (BBU) with a capacity-constrained fronthaul. In the AAS configuration, multiple passive antenna elements and radio units are integrated into a single box. This paper considers precoded downlink transmission over a single-cell MU-MIMO system. We study optimized linear precoding for AAS with a limited-capacity fronthaul, which requires the precoding matrix to be quantized. We propose a new precoding design that is aware of the fronthaul quantization and minimizes the mean-squared error at the receiver side. We compute the precoding matrix using a sphere decoding (SD) approach. We also propose a heuristic low-complexity approach to quantized precoding. This heuristic is computationally efficient enough for massive MIMO systems. The numerical results show that our proposed precoding significantly outperforms quantization-unaware precoding and other previous approaches in terms of the sum rate. The performance loss for our heuristic method compared to quantization-aware precoding is insignificant considering the complexity reduction, which makes the heuristic method feasible for real-time applications. We consider both perfect and imperfect channel state information.
\end{abstract}
\IEEEpeerreviewmaketitle


\section{Introduction} \label{sec:intro}

Multi-user multiple-input and multiple-output (MU-MIMO) is a transmission technique where a base station (BS) with multiple antennas communicates with multiple user equipments (UEs) on the same time-frequency resources. In MU-MIMO systems, spatial diversity and multiplexing techniques are used to obtain reliable communication links and higher data rates, respectively \cite{Swales1990a,Gesbert2007a}. However, multiuser interference induces a performance loss when the BS communicates with multiple UEs and must be controlled. Precoding is the key method that is implemented at the BS to minimize interference in the downlink \cite[Ch.~1]{bjornson2017massive}. 

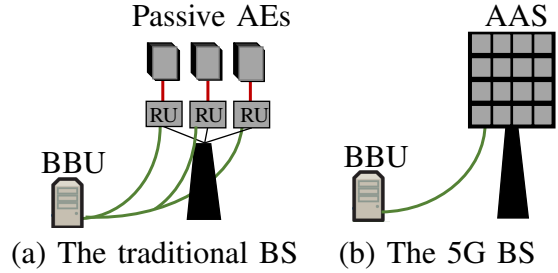
\begin{figure}[t!]
  \centering
   \begin{overpic}[scale=0.5,unit=1mm]{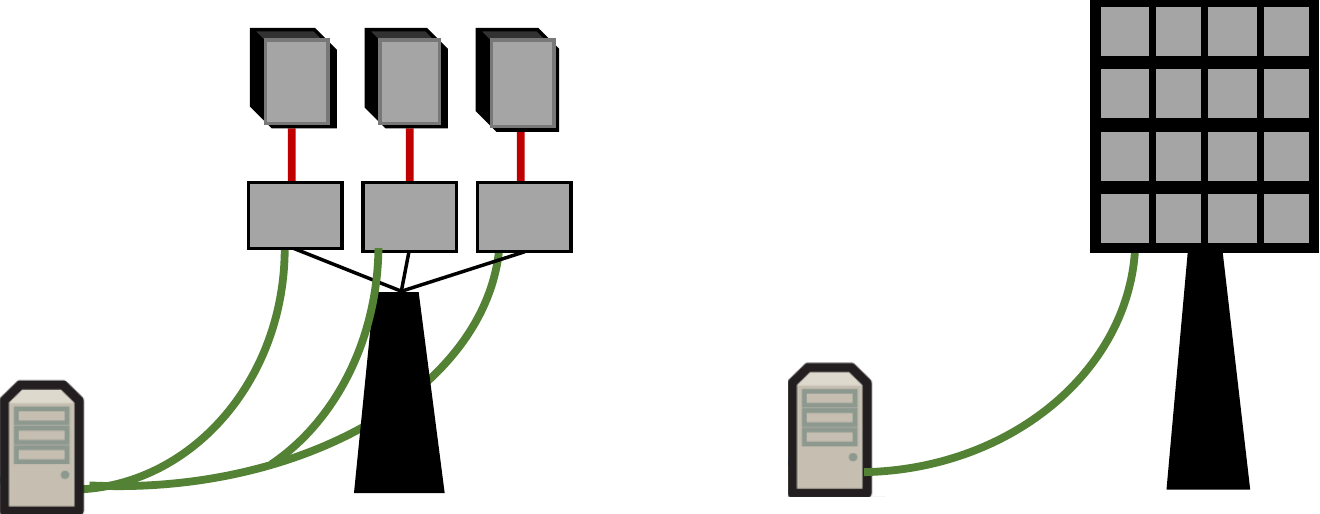}
  \put(-2,11){BBU}%
  \put(58,12){BBU}%
  \put(19.5,21){\scriptsize RU}%
  \put(37.5,21){\scriptsize RU}%
  \put(28.5,21){\scriptsize RU}%
  \put(16,40){Passive AEs}%
  \put(86,40){AAS}%
  \put(-8,-7){(a) The traditional BS}%
  \put(57,-7){(b) The 5G BS}%
   \end{overpic}
   \vspace{5mm}
\caption{Two different BS configurations: (a) A traditional BS with cables between the passive antenna elements (AEs), the radio units (RUs), and the baseband unit (BBU); (b) A 5G BS with a digital fronthaul between the BBU and an advanced antenna system (AAS) with many integrated AEs and RUs. The former case requires many more cables and boxes, while the digital fronthaul between the BBU and AAS is the main limitation in the latter case.}
\label{fig:systemmodel}
\end{figure}

Nowadays, the hardware configuration of BSs is different from the traditional one, which has a significant effect on how the precoded signals are computed. In the traditional configuration, each BS contains one baseband unit (BBU) and then two boxes per antenna: one passive antenna element (AE) and one radio unit (RU), as depicted in Fig.~\ref{fig:systemmodel}(a). Hence, to build a 16-antenna MU-MIMO BS, we need 33 interconnected boxes. However, 5G BSs integrate all AEs and RUs into a single enclosure, called an \emph{advanced antenna system (AAS)} \cite[Ch.~1]{asplund2020advanced} and shown in Fig.~\ref{fig:systemmodel}(b). This hardware evolution has made massive MIMO practically feasible with a compact form factor \cite{Bjornson2019d} and also enables the BBU functions to be virtualized in the cloud through the migration to the centralized radio access network (C-RAN) architecture \cite{peng2015fronthaul}. A new implementation bottleneck in these systems is the digital fronthaul between the AAS and BBU, which needs a capacity that grows with the number of antennas and incurs a finite resolution on all signals. This interface must carry received uplink signals (to be decoded at the BBU) and precoded downlink signals, which are computed at the BBU. This paper focuses on the downlink and proposes a novel fronthaul quantization-aware precoding design.

\subsection{Related Works}
The impact of impairments in analog hardware on the communication performance of MU-MIMO systems has received much attention in prior literature (see e.g., \cite{wenk2010mimo,Zhang2012a,Bjornson2014a,Mollen2018a,aghdam2020distortion}). Moreover, there are related works on quantization distortion caused by low-resolution analog-to-digital converters (ADC) in the uplink \cite{mollen2016uplink,studer2016quantized} and 
low-resolution digital-to-analog converters (DAC) in the downlink \cite{jacobsson2017quantized,mezghani2009transmit,jacobsson2019linear,castaneda2019finite}. 
For example, \cite{jacobsson2019linear} used Bussgang's decomposition to derive a lower bound on the downlink capacity when the precoded signals are sent using low-resolution DACs. A key characteristic of these prior works is that the distortion is created in the RU, i.e., in the analog domain, or in the converters. This implies that the transmit signal obtained after precoding is distorted, which differs from the setup in this paper where only the precoding matrix is distorted. 

Alternatively, symbol-level
precoding techniques, where the transmitted signals are designed based on the knowledge of both channel state information (CSI) and the data symbols have been recently proposed for downlink MU-MIMO systems with low-resolution DACs \cite{tabeshnezhad2021reduced,masouros2010correlation,tsinos2018symbol}. In \cite{tsinos2018symbol} a novel symbol-level precoding technique is developed that supports systems with DACs of any resolution, and it is applicable for any signal constellation. The authors in \cite{tabeshnezhad2021reduced} propose a reduced complexity
precoding method based on linear programming and constructive interference. Different from these non-linear precoding schemes, our linear precoding design is independent of the data and can be used for arbitrary many data symbols, leading to vastly lower complexity.

Another related line of work is quantized feedback for MU-MIMO \cite{Love2008a,jindal2006mimo}, where the UEs estimate and feed back their channels to the BS. The precoding matrix is then computed based on the quantized channels, which leads to extra unremovable interference even if zero-forcing is used \cite{jindal2006mimo}. The key difference from this paper is that the feedback quantization appears before the precoding (the CSI is quantized), instead of after. The effect of limited fronthaul capacity was studied in \cite{park2013joint,simeone2009downlink,park2014fronthaul, parida2018downlink} among others, but the focus was not on precoding design. In \cite{castaneda2019finite}, the authors proposed a finite-resolution linear precoding method for massive MIMO. An all-digital C-RAN deployment was also considered in \cite{huang2018functional}, but for the uplink direction. A sum-rate maximization problem with capacity-constrained fronthaul and optimal bandwidth allocation is discussed in \cite{park2017massive}, where the authors assume that the precoding matrix and data symbols are separately transported over the fronthaul link.

In this paper, we analyze and mitigate the precoding distortion that occurs over the digital interface between the BBU and AAS, where the key difference is that the precoding matrix is quantized before the transmit signal is computed; that is, before the quantized precoding matrix is multiplied with the data symbols at the AAS. We develop one benchmark method for computing such a precoding matrix in an optimized manner and also propose a low-complexity method. The work is focused on block-level precoding, where the same quantized precoding matrix is utilized for all the symbols in a coherence block.

\subsection{Contributions}
We consider MU-MIMO precoding quantization over a limited-capacity fronthaul connection. Since the data symbols originate from a finite-resolution codebook and change much more rapidly than the precoding matrix, an efficient implementation will send these quantities separately over the fronthaul so that only the precoding matrix must be quantized. We formulate and solve a novel quantization-aware precoding problem, where the communication performance after quantization is maximized.
The main contributions are:

\begin{itemize}
    \item Inspired by practical AAS implementation, we formulate a new downlink block-level precoding framework where the precoding matrix is quantized when sent over the limited-capacity fronthaul from the BBU to the AAS, while data symbols require no further quantization.
    
    \item We formulate a quantization-aware linear precoding problem, where the precoding matrix is selected to minimize the mean-squared error (MSE) at the receiver side. Unlike previous works that consider one-bit quantized MU-MIMO \cite{mollen2016uplink,tabeshnezhad2021reduced}, we cast an optimization problem that manages multi-bit quantization. We use mixed-integer convex programming (MICP) \cite{wolsey2007mixed,grant2014cvx,CoeyLubinVielma2018,gurobi2020reference} to solve the problem to global optimality for a fixed precoding coefficient. 
    
    \item To decrease the complexity of solving our mixed-integer problem, we propose a novel sphere precoding (SP) algorithm by using the sphere decoding (SD) technique. We compare the run time of these algorithms to show the benefits for practically-sized problems.
    
    \item We suggest a heuristic approach that begins with quantization-unaware precoding and then updates the precoding columns sequentially. This approach is sufficiently efficient to be used in the massive MIMO case.
    
    \item We provide numerical results to show the benefits of the proposed quantization-aware precoding technique and our heuristic approach over the quantization-unaware baseline and also what is presented in \cite{castaneda2019finite}, in terms of minimizing the MSE. We describe how the number of quantization levels, number of UEs, and CSI quality affect the performance. 
\end{itemize}

This paper refines the problem formulation from the conference version of this work \cite{yasaman2022} and develops an efficient sphere precoding implementation as well as a low-complexity approach that did not appear in \cite{yasaman2022}.

\subsection{Notations}
The set of integer, real, and complex numbers are denoted as $\mathbb{Z}$, $\mathbb{R}$, and $\mathbb{C}$, respectively. Matrices and vectors are represented by upper and lower bold case letters, as $\textbf{X}$ and $\textbf{x}$. The element in the $m$-th row and $k$-th column of a matrix $\textbf{X}$ is denoted by $\textbf{X}[m,k] = x_{m,k}$ while $\textbf{x}[m] = {x}_{m}$ denotes the $m$-th element of  $\textbf{x}$. In addition $\textbf{x}_k$ indicates $k$-th column of the matrix $\textbf{X}$. The identity matrix of size $M \times M$ is denoted as $\textbf{I}_M$. Besides, $\textbf{1}_{M \times K}$ and $\textbf{0}_{M \times K}$ represent matrices of size $M \times K$ in which all elements 
are one and zero, respectively. The notation $| \cdot |$ represents the absolute value, $\Vert \cdot \Vert_2$ denotes the Euclidean norm of a vector, and $\Vert \cdot \Vert_\mathrm{F}$ represents the Frobenius norm of a matrix. The  Kronecker product is denoted by $\otimes$. The transpose, conjugate, conjugate transpose,  matrix inversion, vectorization, and trace are given by $(\cdot)^\mathrm{T}$, $(\cdot)^*$, $(\cdot)^\mathrm{H}$, $(\cdot)^{-1}$, $\mathrm{vec}(\cdot)$, and $\mathrm{tr}(\cdot)$, respectively. Furthermore, $\mathfrak{R} \{ \cdot \}$ denotes the real part and $\mathfrak{I} \{ \cdot \}$ denotes the imaginary part of a complex number. $\mathbb{E}[X]$ represents the statistical average of a random variable $X$, and $X\sim \mathcal{CN}(0,\sigma^2)$ denotes a circularly symmetric complex Gaussian distribution with variance $\sigma^2$. 



\subsection{Paper Outline}
The organization of the paper is as follows. Section \ref{sec2} presents an overview of the proposed system model and discusses quantization-unaware precoding as a baseline. Section \ref{sec3solve} describes the proposed precoding framework, which that is called the quantization-aware precoding approach, and the SP approach that is used to solve the problem. In Section \ref{sec4sumrate}, we propose a heuristic approach with vastly reduced complexity. Section \ref{sec5num} provides numerical results and  evaluates the performance in terms of sum rate for the new schemes and existing benchmarks. Finally, Section \ref{sec6con} summarizes the main conclusions and provides suggestions for future work.
\section{System Model} \label{sec2}

We consider the downlink of a single-cell MU-MIMO system, where a BS equipped with an AAS with $M$ antenna-integrated radios serves $K$ single-antenna UEs on the same time-frequency resource. The AAS is connected to a BBU through a digital fronthaul link with limited capacity. Hence, any signal that is sent over the fronthaul must be quantized to finite resolution. Each transmitted signal vector is the product between a precoding matrix and a vector with data symbols, wherae the former is assumed fixed for the duration of the transmission while the latter changes at the symbol rate. 
The BBU encodes the data and computes a precoding matrix based on CSI and then forwards it to the AAS. As the data symbols represent bit sequences from a channel coding codebook, we can send them over the fronthaul  without quantization errors and map them to modulation symbols at the AAS. However, the precoding matrix normally contains arbitrary complex-valued entries and must be quantized before being sent over the limited-capacity fronthaul. The quantized precoding matrix is then multiplied with the UEs' data symbols at the AAS, and finally the product is transmitted wirelessly. 

\begin{remark}
The fronthaul signaling can be vastly reduced by conveying the precoding matrix and signal vector separately. To demonstrate this, suppose $N_{\mathrm{precoder}}$ is the number of quantization bits per real dimension and $\tau$ is the number of signal vectors per transmission block. The required fronthaul capacity is $C_{\mathrm{separate}} = 2MKN_{\mathrm{precoder}}+\tau K \mathsf{SE}$ bits per block, where the spectral efficiency $\mathsf{SE}$ represent the number of bits per data symbol.
This should be compared against multiplying the precoding matrix and signal vector at the BBU and then sending the resulting $M$-length vectors over the fronthaul.
This joint fronthaul signaling requires 
 $C_{\mathrm{joint}}= M\tau N_{\mathrm{res}} \mathsf{SE}$ bits per block, where the factor $N_{\mathrm{res}}$ determines how much larger the quantization resolution per entry is compared to the SE of the data symbols.
For example, consider $16$-QAM with $\mathsf{SE}=4$ bit/symbol, $M=16$ antennas, $K=4$ UEs, $\tau = 200$ symbols per block, $N_{\mathrm{precoder}}=3$ bits per real dimension, and $N_{\mathrm{res}}=3$ times higher quantization resolution than spectral efficiency.
We then get
$C_{\mathrm{separate}} =  3584$ while $C_{\mathrm{joint}}= 38400$, which shows that the fronthaul load is reduced by $10$ times when following the proposed approach of transmitting the precoding matrix and signal vector separately. This is why this paper only considers the former option.
\end{remark}

Before analyzing the proposed quantization-aware  precoding scheme in Section \ref{sec3solve}, we first introduce the considered transmission model in the following subsections. We also discuss the model of imperfect CSI that will be used in parts of this paper. Then, the conventional uniform quantizer-mapping and the quantization operator are defined.

\subsection{Downlink transmission}\label{sec2chamodel}

To focus on algorithmic development, we assume that the BBU has perfect CSI and neglect all potential transceiver hardware impairments. Later in this paper, we also demonstrate how the algorithms can be applied along with imperfect CSI. The downlink system model can be written as
\begin{equation}
\textbf{y}=\textbf{H}\textbf{x}+ \textbf{n},
\end{equation}
where $\textbf{y} = [y_1, \ldots, y_K]^\mathrm{T} \in \mathbb{C}^{K} $ contains the received signals at all the UEs and $y_k \in \mathbb{C}$ denotes the signal received at the $k$-th UE. The downlink channel matrix $\textbf{H} \in \mathbb{C}^{K \times M}$ has entries  $h_{k,m}$ for $k=1, \ldots ,K$ and $m=1, \ldots ,M$. It represents a narrowband channel and might be one subcarrier of a multi-carrier system. In the latter case, the algorithms developed in this paper can be applied individually on each subcarrier. The vector $\textbf{n} = [n_1, \ldots, n_K]^\mathrm{T} \in \mathbb{C}^{K}$ represents additive white noise, where has independent and identically distributed (i.i.d.) entries $n_k \sim \mathcal{CN}(0,N_0)$. The precoded signal vector is denoted by $\textbf{x} = \textbf{P}\textbf{s}$, where the vector $\textbf{s} = [s_1, \ldots, s_K]^\mathrm{T}\in \mathcal{O}^{K}$ contains the data symbols and $s_k$ denotes the random data symbol intended for UE $k$ normalized to unit power. Here, $\mathcal{O}$ is the finite set of constellation points (e.g., a conventional QAM alphabet). 
We consider \textit{linear block-level precoding} where $\textbf{P}$ depends on the channel matrix $\textbf{H}$, but not on the data symbols.

The precoding matrix $\textbf{P}$ is obtained from the BBU and has entries  $p_{m,k} \in \mathcal{P}$ for $k=1, \ldots ,K$ and $m=1, \ldots ,M$, where the fronthaul quantization alphabet set $\mathcal{P}$ is defined as
\begin{equation}
   \mathcal{P} = \{ l_{R} + jl_{I} : l_{R},l_{I} \in \mathcal{L}  \}.
   \label{eq:quantizationset}
\end{equation}
We assume that the same quantization alphabet is used for the real and imaginary parts. Here $\mathcal{L}= \{ l_0, \ldots ,l_{L-1} \}$ contains the set of real-valued quantization labels, $L=|\mathcal{L}|$ denotes the number of quantization levels and 
$N=\log_2 (L)$ is the number of quantization bits per real dimension. Note that $\mathcal{P}$ coincides with the complex numbers set $\mathbb{C}$ in the case of infinite resolution. 

The AAS computes the precoded signal vector $\textbf{x} = \textbf{P}\textbf{s}$ and it must satisfy the following average power constraint:
\begin{equation}
    \mathbb{E}[\left \Vert \textbf{x} \right \|_2^2]
    = \left \Vert \textbf{P} \right \|_{\mathrm{F}}^2 \le  q, \label{eq:powercons}
\end{equation}
where $\left \Vert \textbf{P} \right \|_{\mathrm{F}}$ denotes the Frobenius norm and $q$ is the maximum average transmit power of the downlink signals.
The equality in \eqref{eq:powercons} follows from the fact that the data symbols are i.i.d. and have unit power.\footnote{Note that all expectations are taken as  if the channel matrix $\textbf{H}$ is deterministic.
In the numerical results, we will average over the results obtained using different random realizations of $\textbf{H}$. In those cases, the expectations are instead conditioned on a single channel matrix realization, since the precoding is a function only of the current channel realization.}

The $k$-th UE needs to estimate the transmitted data symbol $\hat{s}_k \in \mathbb{C}$ based on its received signal $y_k$. We assume that the UE computes the estimate as $\hat{s}_k = \beta_k y_k$, where the linear equalization is based on the precoding factor $\beta_k\in \mathbb{C}$ that is selected to minimize the MSE $\mathbb{E}[  \left | {s}_k -  \hat{{s}}_k \right |^2]$. In the precoding design, we will treat all UEs equally by minimizing their sum MSE
\begin{equation}
        \mathbb{E} \left[  \left \Vert \textbf{s} -  \hat{\textbf{s}} \right \|_2^2 \right] = \mathbb{E} \left[  \left \Vert \textbf{s} -  \textbf{B} \textbf{y} \right \|_2^2 \right],
        \label{eq:mmseform}
\end{equation}
where $\hat{\textbf{s}} =[\hat{s}_1, \hat{s}_2, \ldots, \hat{s}_K]^\mathrm{T}$ is a vector containing the estimated data symbols and $\textbf{B}$ is a $K \times K$ diagonal matrix with  $\beta_k$ as the $k$-th diagonal element. We also define the vector $\boldsymbol \beta = [\beta_1, \beta_2, \ldots, \beta_K]^\mathrm{T}$ with the precoding factors, because it will later be treated as an optimization variable.

\subsection{Imperfect CSI}

The BBU uses its available CSI to select the downlink precoding matrix. 
The CSI can be represented by the channel matrix estimate $\hat{\textbf{H}} \in \mathbb{C}^{K \times M}$ with entries  $\hat{h}_{k,m} $.
As the main focus of this work is on the fronthaul quantization effects, we will primarily consider perfect CSI, i.e., $\hat{\textbf{H}} = \textbf{H}$. However, the same algorithms can be used if the BBU has imperfect CSI, but evaluates the sum MSE in \eqref{eq:mmseform} as if the CSI is perfect. That is a common way of doing precoding with imperfect CSI \cite{Love2008a,jindal2006mimo}. 

We will take imperfect CSI into account in our numerical results and we will motivate the corresponding modeling of CSI imperfections in this section. Suppose all the channel coefficients $h_{k,m}$ for $k=1, \ldots ,K$ and $m=1, \ldots ,M$ are modeled as i.i.d. complex Gaussian random variables with variance $\gamma_k$, i.e.,
$h_{k,m} \sim \mathcal{CN}(0,\gamma_k)$. All the links are assumed to be block fading, and the channel parameters are constant within one block and change independently from block to block. 
We consider a time-division duplex (TDD) scenario, where the uplink channels are equal to the downlink channels due to reciprocity, and make use of the typical estimation framework from the massive MIMO literature
\cite{bjornson2017massive}.
Each UE transmits a known \textit{uplink pilot} signal $x_k \in \mathbb{C}$ with power $\Bar{q}_k$. The received signal vector from the $k$-th UE at the BBU is given by
\begin{equation}
    \Bar{\textbf{y}}_k={\textbf{h}}_kx_k+\Bar{\textbf{n}}_k,
\end{equation}
where $\Bar{\textbf{y}}_k = [\Bar{y}_{k,1}, \Bar{y}_{k,2}, \ldots, \Bar{y}_{k,M}]^\mathrm{T}$, ${\textbf{h}}_k = [h_{k,1}, h_{k,2}, \ldots, h_{k,M}]^\mathrm{T}$,
 and $ \Bar{\textbf{n}}_k = [\Bar{n}_{k,1}, \Bar{n}_{k,2}, \ldots, \Bar{n}_{k,M}]^\mathrm{T} \sim \mathcal{CN}(0,N_0\textbf{I}_M )$ is independent noise. The MMSE estimate of $h_{k,m}$ is \cite[Ch.~3]{bjornson2017massive}
\begin{align}
    \hat{h}_{k,m} = \frac{\sqrt{\Bar{q}_k} \gamma_k}{N_0+\Bar{q}_k \gamma_k} \bar{y}_{k,m} \sim \mathcal{CN} \left(0,\frac{\Bar{q}_k \gamma_k^2}{N_0+\Bar{q}_k \gamma_k} \right).
    \label{eq:channel estimate}
\end{align}
We notice that this model only considers the estimation errors that occur due to the limited uplink SNR, while we have neglected further fronthaul quantization errors.

\subsection{Quantizer Function} \label{sec2quantize}

The limited-capacity fronthaul is modeled as a quantizer. Since uniform quantization is often used in practice, we model our quantizer function $\mathcal{Q}(\cdot) : \mathbb{C} \to \mathcal{P}$ as a symmetric uniform quantization with step size $\Delta$. Each entry of the quantization labels $\mathcal{L}$ is defined as 
\begin{equation}
    l_z  = \Delta  \left( z- \frac{L-1}{2}  \right), \quad z=0, \ldots, L-1.
\end{equation}
Furthermore, we let $\mathcal{T} = \{ \tau_0, \ldots, \tau_L \}$, where $-\infty = \tau_0 < \tau_1 < \ldots < \tau_{(L-1)} < \tau_{L} =\infty$, specify the set of the $L + 1$ quantization thresholds. For uniform quantizers, the thresholds
are 
\begin{equation}
    \tau_z =  \Delta  \left( z- \frac{L}{2}  \right), \quad z=1, \ldots, L-1.
\end{equation}
The quantizer function $\mathcal{Q}(\cdot)$ can be uniquely described by the set of quantization labels $\mathcal{L} = \{ l_z : z=0, \ldots, L-1 \}$ and the set of quantization thresholds $\mathcal{T}$. The quantizer maps an input $r \in \mathbb{C} $ to the quantized output $\mathcal{Q}(r) = l_o + jl_l \in \mathcal{P}$, where the set is defined in \eqref{eq:quantizationset}, if
$\mathfrak{R}\{ \mathcal{Q}(r) \} \in  [\tau_o,\tau_{o+1})$ and $\mathfrak{I}\{ \mathcal{Q}(r) \} \in  [\tau_l,\tau_{l+1})$.
The step size $\Delta$ of the quantizer should be chosen to minimize the distortion between the quantized output and unquantized input. The optimal step size $\Delta$ depends on the statistical distribution of the input, which in our case depends on the precoding scheme and  the channel model. Since the distribution of the precoding matrix elements is generally unknown and varying with the user population, we set the step size to minimize the distortion under the maximum-entropy assumption that the per-antenna input to the quantizers is $\mathcal{CN}(0,\frac{q}{KM})$ distributed, where the variance is selected so that the sum power of the elements matches with the power constraint in \eqref{eq:powercons}. The corresponding optimal step size for the normal distribution was found in \cite{hui2001unifquantized}.

\subsection{Quantization-unaware precoding} \label{sec2unaware}

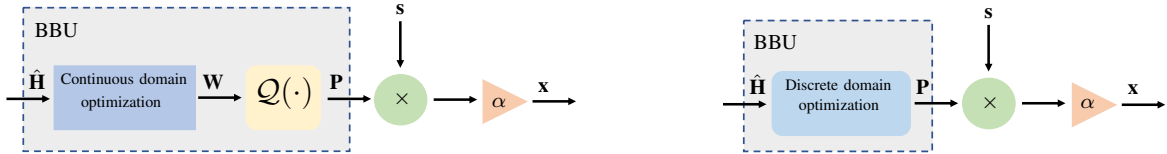
\begin{figure}[!t]
    \centering

  \begin{overpic}[scale=0.43,unit=1mm]{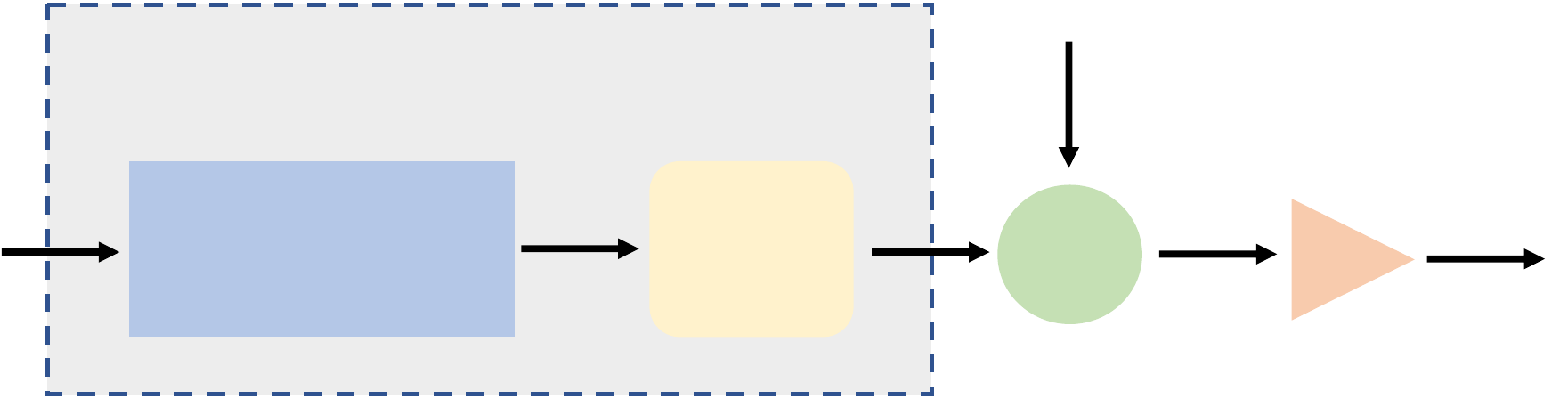}
  \put(4,11){\scriptsize$\hat{\textbf{H}}$}%
  \put(5,20){\scriptsize BBU}%
  \put(9.5,12){\tiny Continuous domain } 
  \put(13,8){\tiny  optimization} 
  \put(34,11){\scriptsize${\textbf{W}}$}%
  \put(43,9){$\mathcal{Q}(\cdot)$} 
  \put(56,11){\scriptsize${\textbf{P}}$}  \put(67.5,24){\scriptsize${\textbf{s}}$}%
  \put(67,8.5){\scriptsize$\times$}
  \put(92,10.5){\scriptsize${\textbf{x}}$} 
  \put(84,8){\scriptsize${\alpha}$} 
  \put(-6,-9){\small(a) Quantization-unaware precoding}%
  \end{overpic}
  \qquad \qquad
\begin{overpic}[scale=0.43,unit=1mm]{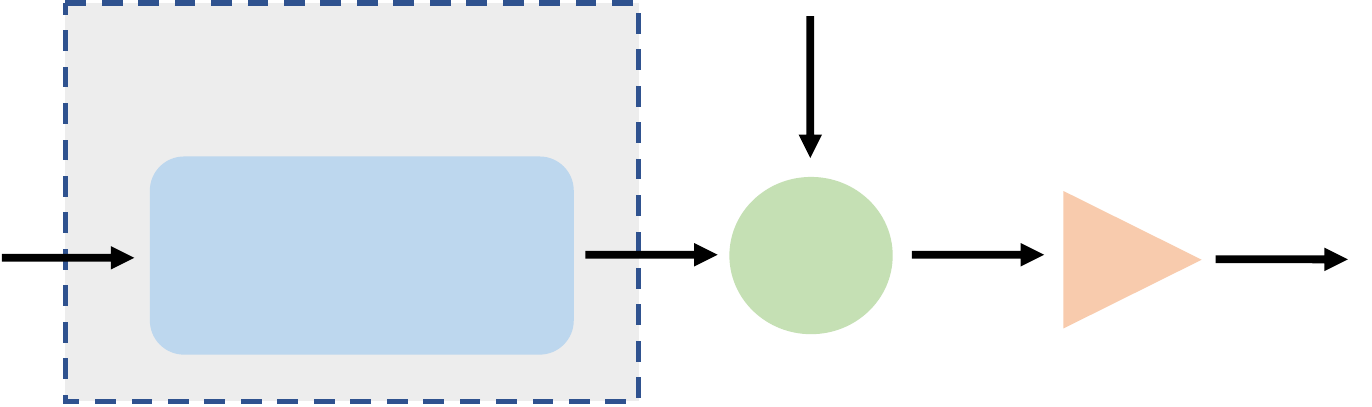}
  \put(6,13){\scriptsize$\hat{\textbf{H}}$}%
  \put(7,23){\scriptsize BBU}%
  \put(43,13){\scriptsize${\textbf{P}}$}%
  \put(58.2,30){\scriptsize${\textbf{s}}$}%
  \put(57.5,10){\scriptsize$\times$}  
  \put(90,13){\scriptsize${\textbf{x}}$}   \put(79.5,10){\scriptsize$\alpha$}      \put(14,14){\tiny Discrete domain } 
  \put(17,9.5){\tiny  optimization} 
  \put(-10,-12){\small(b) Quantization-aware precoding}%
   \end{overpic}
   \vspace{5mm}
   \caption{There are several key differences between the baseline quantization-unaware precoding in (a) and our proposed quantization-aware precoding scheme in (b).}
   \label{fig:difference}
\end{figure}

When it comes to quantized precoding, we will consider two cases. The naive \textit{baseline} approach is to first design a precoding matrix $\textbf{W} \in \mathbb{C}^{M\times K}$ based on the available CSI and any precoding scheme designed with infinite resolution and then quantize using $\mathcal{Q}(\cdot): \mathbb{C}^{M \times K} \to \mathcal{P}^{M \times K}$ so that its output can be sent over the fronthaul. The BBU can use any precoding scheme from the literature but should make sure that $\|\textbf{W}\|_F^2 = q$, so that it would satisfy the power constraint in (\ref{eq:powercons}) and thereby fit the dynamic range of the quantizer. We will refer to this as \emph{quantization-unaware precoding}. 
As depicted in Fig.~\ref{fig:difference}(a), the quantization-unaware precoding matrix is obtained as
\begin{equation}
    \textbf{P} = \mathcal{Q}(\textbf{W}),
    \label{eq:unaware}
\end{equation}
where all the calculations are done at the BBU. 
The quantized precoding matrix $\textbf{P}$ and the data symbols vector $\textbf{s}$ are sent separately over the fronthaul.
The precoded signal vector $\textbf{x}$ is calculated at the AAS as
\begin{equation}
    \textbf{x} = \alpha \textbf{P} \textbf{s}
\end{equation}
where the scaling factor $\alpha = \sqrt{q/ \Vert \textbf{P} \|_{\mathrm{F}}^2}$ is computed at the AAS with the available information. 
This scaling factor makes sure that maximum power is used during transmission and is needed since the condition $\left \Vert \textbf{W} \right \|_{\mathrm{F}}^2 =  q$ does not imply $\left \Vert \textbf{P} \right \|_{\mathrm{F}}^2 =  q$.

It is worth mentioning that there are prior works (e.g., \cite{jacobsson2017quantized}) that consider a seemingly similar quantization-aware procedure but quantize the precoded signal $\textbf{x}$ and not just the precoding matrix. This is relevant when the quantization distortion comes from the DAC but not for our setup. 

As we mentioned, the BBU can use any \textit{unnormalized} precoding matrix $\textbf{W}$ and then quantize it. Some classical schemes to compute \textit{continuous} precoding matrices are transmit Wiener filtering (WF), zero-forcing (ZF), and maximum ratio transmission (MRT) \cite{joham2005linear,bjornson2014beamforming}. The WF precoding scheme is the most desirable one and is derived in \cite{joham2005linear} by minimizing the  MSE in \eqref{eq:mmseform}. The corresponding unnormalized continuous precoding matrix ${\textbf{W}^\mathrm{WF}}$ is computed as
\begin{equation}
    \textbf{W}^\mathrm{WF}= \textbf{H}^\mathrm{H} \left(\textbf{H}\textbf{H}^\mathrm{H}+\frac{KN_0}{q}\textbf{I}_K \right)^{-1}.
    \label{eq:wf}
\end{equation}

Instead of following this two-step procedure, where a continuous precoding matrix is first computed and then quantized to fit the discrete fronthaul quantization alphabet, we could craft an algorithm that directly computes a precoding matrix where the elements come from this discrete alphabet.
This alternative is illustrated in Fig.~\ref{fig:difference}(b) and will be called \emph{quantization-aware precoding}. In Section \ref{sec3solve}, we develop an optimization algorithm that minimizes the MSE, similarly to WF precoding, by taking knowledge of the quantization alphabet into account.

\section{Quantization-aware Precoding}\label{sec3solve}

The uncontrolled quantization effect in the quantization-unaware precoding scheme leads to extra interference and reduced beamforming gain.  Thus, it is not the optimal scheme.
For example, canceling all interference using zero-forcing is optimal when the signal-to-noise ratio (SNR) is high and there is no quantization \cite{joham2005linear,bjornson2014beamforming}, while it will not be the case in our setup due to the fronthaul quantization. In this section, we propose a scheme that finds an optimal quantization-aware precoding
that minimizes the MSE between the received signal and the transmitted symbol vector $\textbf{s}$ under the power constraint in \eqref{eq:powercons}.
\subsection{Optimal Quantization-Aware Precoding}
We formulate the quantization-aware precoding optimization problem as

\begin{mini}|l|
	  {\textbf{P}\in \mathcal{P}^{M \times K} , \boldsymbol \beta \in \mathbb{C}^K }{\mathbb{E}[  \left \Vert \textbf{s} -  \textbf{B} \textbf{y}\right \|^2]}{}{}
	  \addConstraint{\left \Vert \textbf{P} \right \|_{\mathrm{F}}^2\le q.} \label{eq:mmse}
\end{mini}
There are two optimization variables in \eqref{eq:mmse}: the precoding matrix $\textbf{P}$ and the vector $\boldsymbol \beta$ containing precoding factors (that are used at the receiver side) that are the elements on the diagonal of $\textbf{B}$.

A Lagrangian multiplier approach can be taken to reformulate the optimization problem in \eqref{eq:mmse}. A simplified Lagrangian function can be written as follows

\begin{equation}
    \mathfrak{L}(\textbf{P}, \boldsymbol \beta, \lambda) = \mathrm{tr} \Big( \textbf{I}_K-\textbf{B} \textbf{H}\textbf{P}-  \textbf{P}^\mathrm{H}
 \textbf{H}^\mathrm{H}\textbf{B}^\mathrm{H} + \textbf{B}\textbf{H}\textbf{P}\textbf{P}^\mathrm{H}\textbf{H}^\mathrm{H}\textbf{B}^\mathrm{H} \Big)  + \lambda \big( \mathrm{tr}(\textbf{P}\textbf{P}^\mathrm{H})-q \big).
\end{equation}
In the next subsection, we present a step-by-step guide to simplifying the Lagrangian function $\mathfrak{L}$.  Then, for a given quantized precoding matrix $\textbf{P}$,
we can compute the optimal value of $\beta_k$ by taking the Wirtinger derivative $\frac{\partial \mathfrak{L}}{\partial \beta^*} $ and equate it to zero. This leads to the equation
\begin{equation}
    \beta_k [\textbf{P}^\mathrm{H}\textbf{H}^\mathrm{H}\textbf{H}\textbf{P}]_{k,k}  + \beta_k N_0 = [\textbf{P}^\mathrm{H}\textbf{H}^\mathrm{H}]_{k,k} \ ,
\end{equation}
and we obtain the optimal precoding factor as 
\begin{equation}
    \beta_k^{\mathrm{Opt}} = \frac{[\textbf{P}^\mathrm{H}\textbf{H}^\mathrm{H}]_{k,k}}{[\textbf{P}^\mathrm{H}\textbf{H}^\mathrm{H}\textbf{H}\textbf{P}]_{k,k} + N_0}\ .\label{eq:betaopt}
\end{equation}
If we substitute (\ref{eq:betaopt}) into (\ref{eq:mmse}), the remaining combinatorial problem of finding $\textbf{P}$ will be computationally intractable. An iterative method is possible where we switch between optimizing $\textbf{P}$ and $\boldsymbol \beta$ while keeping the other one fixed, but we have noticed experimentally that $\boldsymbol \beta$ will not change much between iterations. Hence, we propose to pick a judicious value of $\boldsymbol\beta$ and then solve the remaining problem
\begin{mini}|l|
	  {\textbf{P}\in \mathcal{P}^{M \times K} }{\mathbb{E}[  \left \Vert \textbf{s} -  \textbf{B}^{\mathrm{Opt}} \textbf{y}\right \|^2]}{}{}
	  \addConstraint{\left \Vert \textbf{P} \right \|_{\mathrm{F}}^2\le q,} \label{eq:mmsebetafixed}
\end{mini}
where $\textbf{B}^{\mathrm{Opt}}$ is the diagonal matrix that has the optimal precoding factor 
$\beta_k^{\mathrm{Opt}}$ as its $k$-th diagonal element. In the numerical evaluation in Section \ref{sec5num}, we will use $\beta_k = \beta_k^{\mathrm{WF}}$, which is calculated by substituting \eqref{eq:wf} into \eqref{eq:betaopt} and is computed for WF precoding without fronthaul quantization (i.e., infinite resolution). The expression in \eqref{eq:betaopt} can be simplified for WF precoding factors into
\begin{equation}
\beta_k^{\mathrm{WF}}=\frac{1}{\sqrt{q}} \Big[ \Big( \textbf{H}^\mathrm{H} \Big(\textbf{H}\textbf{H}^\mathrm{H}+\frac{KN_0}{q}\textbf{I}_K\Big)^{-2}\textbf{H} \Big)\Big]_{k,k}^{1/2} \  .\label{eq:betawf}
\end{equation}
\subsection{A Mixed-Integer Programming Reformulation}

We will solve \eqref{eq:mmsebetafixed} by reformulating the problem in a way that enables the use of MICP methods. To find the MSE-minimizing precoding matrix $\text{P}$ for an arbitrary fixed diagonal matrix $\textbf{B}$, first we simplify the objective function of the problem (\ref{eq:mmsebetafixed}) as 

\begin{align}
\mathbb{E}[  \left \Vert \textbf{s} -  \textbf{B} \textbf{y}\right \|^2] & = \mathbb{E}[  \left \Vert \textbf{s} -  \textbf{B} \textbf{H}\textbf{P}\textbf{s} -  \textbf{B} \textbf{n} \right \|^2]  \nonumber
\\  &= \mathrm{tr} \Big( (\textbf{I}_K-\textbf{B} \textbf{H}\textbf{P}) \mathbb{E}[\textbf{s}\textbf{s}^\mathrm{H}] (\textbf{I}_K-\textbf{B} \textbf{H}\textbf{P})^\mathrm{H}   +\textbf{B} \textbf{B}^\mathrm{H} \mathbb{E}[ \textbf{n}\textbf{n}^\mathrm{H} ] \Big) \nonumber
\\ & = \mathrm{tr} \Big( \textbf{I}_K-\textbf{B} \textbf{H}\textbf{P}-  \textbf{P}^\mathrm{H}
 \textbf{H}^\mathrm{H}\textbf{B}^\mathrm{H} + \textbf{B}\textbf{H}\textbf{P}\textbf{P}^\mathrm{H}\textbf{H}^\mathrm{H}\textbf{B}^\mathrm{H} \Big)  + K N_0 \sum_{k=1}^K|\beta_k|^2  \nonumber
 \\ &= \mathrm{tr} \Big(\textbf{P}^\mathrm{H}\textbf{H}^\mathrm{H}\textbf{B}^\mathrm{H}\textbf{B}\textbf{H}\textbf{P} -\textbf{B} \textbf{H}\textbf{P}-  \textbf{P}^\mathrm{H}
 \textbf{H}^\mathrm{H} \textbf{B}^\mathrm{H} \Big)+ K(N_0\sum_{k=1}^K|\beta_k|^2 +1). \label{eq:simplified}
\end{align} 
When minimizing (\ref{eq:simplified}) with respect to $\textbf{P}$, we can drop the constant term $K(N_0\sum_{k=1}^K|\beta_k|^2 +1)$. Hence, the optimal solution to \eqref{eq:mmsebetafixed} coincides with optimal solution to 
\begin{mini}|l|
	  {\textbf{P}\in \mathcal{P}^{M \times K}  }{\mathrm{tr}\Big(\textbf{P}^\mathrm{H}\textbf{H}^\mathrm{H}\textbf{B}^\mathrm{H} \textbf{B}\textbf{H}\textbf{P}- \textbf{B}\textbf{H}\textbf{P} - (\textbf{B}\textbf{H}\textbf{P})^\mathrm{H} \Big)}{}{}
	  \addConstraint{\mathrm{tr}(\textbf{P}\textbf{P}^\mathrm{H}) \le q.} \label{eq:prob}
\end{mini}
To turn (\ref{eq:prob}) into a more tractable vector optimization problem, we define $\textbf{a} = \mathrm{vec}(\textbf{P})$, and $ \textbf{h} = \mathrm{vec}\Big( (\textbf{BH})^\mathrm{T}\Big)$, so we have
\begin{mini}|l|
	  {\textbf{a}\in \mathcal{P}^{MK \times 1}  }{\textbf{a}^\mathrm{H} \left ( \textbf{I}_K \otimes \textbf{H}^\mathrm{H}\textbf{B}^\mathrm{H} \textbf{B}\textbf{H} \right )\textbf{a} -\textbf{h}^\mathrm{T}\textbf{a} - \left ( \textbf{h}^\mathrm{T}\textbf{a}\right )^\mathrm{H} }{}{}
	  \addConstraint{\textbf{a}^\mathrm{H}\textbf{a} \le q,} \label{eq:modify}
\end{mini}
where $\otimes$ denotes the Kronecker product. Due to the quantized search domain $\mathcal{P}^{MK \times 1} $, problem \eqref{eq:modify} resembles a MICP but with complex numbers. We can transfer the problem into an equivalent real-valued form by utilizing the following definitions:
\begin{align}
 \textbf{a}_\mathbb{R} & =
\begin{bmatrix}
\mathfrak{R}\{\textbf{a}\}\\
\mathfrak{I}\{\textbf{a}\}
\end{bmatrix}, \quad
\textbf{c}_\mathbb{R}=
\begin{bmatrix}
\mathfrak{R}\{\textbf{h}\}\\
 \mathfrak{I}\{\textbf{h}\} \nonumber
\end{bmatrix}, \quad \text{and } \\
\textbf{V}_\mathbb{R}&=
\begin{bmatrix}
\mathfrak{R}\{\textbf{I}_K \otimes \textbf{H}^\mathrm{H}\textbf{B}^\mathrm{H} \textbf{B}\textbf{H}\} & -\mathfrak{I}\{\textbf{I}_K \otimes \textbf{H}^\mathrm{H}\textbf{B}^\mathrm{H} \textbf{B}\textbf{H}\} \\
\mathfrak{I}\{\textbf{I}_K \otimes \textbf{H}^\mathrm{H}\textbf{B}^\mathrm{H} \textbf{B}\textbf{H}\} & \mathfrak{R}\{\textbf{I}_K \otimes \textbf{H}^\mathrm{H}\textbf{B}^\mathrm{H} \textbf{B}\textbf{H}\}
\end{bmatrix}.
\end{align}
These definitions enable us to rewrite (\ref{eq:modify}) as 
\begin{mini}|l|
	  {\textbf{a}_\mathbb{R}\in \mathcal{L}^{2MK \times 1} }{\textbf{a}^\mathrm{T}_\mathbb{R} \textbf{V}_\mathbb{R}\textbf{a}_\mathbb{R} -2\textbf{c}^\mathrm{T}_\mathbb{R}\textbf{a}_\mathbb{R} }{}{}
	  \addConstraint{\textbf{a}^\mathrm{T}_\mathbb{R}\textbf{a}_\mathbb{R} \le q ,} \label{eq:real}
\end{mini}
where $\mathcal{L}$ assures that we are not using more quantization steps than allowed.
Both the objective function and constraint of (\ref{eq:real}) are convex functions of $\textbf{a}_\mathbb{R}$. 
However, the search domain of the problem is discrete due to the $\textbf{a}_\mathbb{R}\in \mathcal{L}^{2MK \times 1}$ criteria, thus making this a \emph{mixed-integer convex programming (MICP)} problem. 
There are many numerical general-purpose algorithms for solving such discrete convex optimization problems to global optimality, e.g., see \cite{wolsey2007mixed,grant2014cvx,CoeyLubinVielma2018}. Hence, we can use standard MICP solvers to find the optimal solutions efficiently by defining the following equivalent problem:
\begin{mini}|l|
	  {\textbf{x} \in \mathbb{Z}^{2MK \times 1} }{\textbf{a}^\mathrm{T}_\mathbb{R} \textbf{V}_\mathbb{R}\textbf{a}_\mathbb{R} -2\textbf{c}^\mathrm{T}_\mathbb{R}\textbf{a}_\mathbb{R}   }{}{}
	  \addConstraint{\textbf{a}^\mathrm{T}_\mathbb{R}\textbf{a}_\mathbb{R} \le q}
	  \addConstraint{\textbf{a}_\mathbb{R}=\Delta \left( \textbf{x} - \left(\frac{L-1}{2} \right) \textbf{1}_{2MK \times 1} \right)}
	  \addConstraint{\textbf{0}_{2MK \times 1} \le \textbf{x} \le (L-1)\textbf{1}_{2MK \times 1}.} \label{eq:realinteger}
\end{mini}
In the numerical evaluation in Section \ref{sec5num},  we will use CVX \cite{grant2014cvx} along with the Gurobi solver \cite{gurobi2020reference} to solve this problem. The computational complexity of solving the problem increases exponentially with $MK$, but the numerical results show that the problem is still solvable for some practically-sized MU-MIMO systems. The complexity of problem \eqref{eq:realinteger} is polynomial in the number of quantization levels $L$, but we consider a fixed and relatively small number of quantization bits.

\subsection{Efficient Sphere Precoding Implementation}
\label{subsec:SP}

Solving \eqref{eq:realinteger} with a general-propose solver, such as Gurobi, leads to higher complexity than crafting a dedicated solver for our particular problem. 
In this subsection, we design a more efficient solution that pushes the limits on how large problems can be solved in a real-time application.
To this end, we first notice that problem \eqref{eq:mmsebetafixed} is a so-called integer least-squares problem, where the search space is a finite subset of the infinite lattice. A technique that has previously been proposed as an efficient algorithm to solve closest lattice
point (CLP) problems in a Euclidean sense is called sphere decoding (SD). SD has such lower average computational complexity than a naive exhaustive search \cite{hassibi2005sphere,murugan2006unified,jalden2005complexity,agrell2002closest}. The basic principle of SD is to reduce the number of search points of the skewed lattice that lie within a hypersphere
of radius $d$, which can speed up the process of finding the solution without loss of optimality. One can transform the original
CLP problem into a tree-search problem and then perform a depth-first branch-and-bound procedure and prune branches that exceed the radius constraint to reduce the number of candidate vectors. In this paper, we consider the Schnorr-Euchner SD (SESD) algorithm form \cite{agrell2002closest}. The SESD enumeration sorts candidate symbols in a zig-zag manner. It optimizes the SD algorithm by first checking the smallest child node of the parent node in each layer. The first found feasible solution is often quite good, thus, many branches can be pruned and the calculation complexity can be further lowered.

To adapt our problem \eqref{eq:prob} to match with the form needed by SD algorithms, we proceed as follows. First, we rewrite the objective
function of \eqref{eq:prob} using the Lagrange multiplier $\lambda$ as 
\begin{mini}|l| 
	  {\textbf{P}\in \mathcal{P}^{M \times K}, \lambda \ge 0  }{\mathrm{tr}\Big(\textbf{P}^\mathrm{H}\textbf{H}^\mathrm{H}\textbf{B}^\mathrm{H} \textbf{B}\textbf{H}\textbf{P}- \textbf{B}\textbf{H}\textbf{P} - (\textbf{B}\textbf{H}\textbf{P})^\mathrm{H} \Big) 
	  + \lambda \Big(\mathrm{tr}(\textbf{P}\textbf{P}^\mathrm{H})-q \Big).}{}{}
	  \label{eq:equalsd}
\end{mini}
We can  rewrite the problem as
\begin{mini}|l|
	  {\textbf{P}\in \mathcal{P}^{M \times K}, \lambda \ge 0  }{\mathrm{tr}\Big(\textbf{P}^\mathrm{H}\big(\textbf{H}^\mathrm{H}\textbf{B}^\mathrm{H} \textbf{B}\textbf{H} + \lambda\textbf{I}_M \big)\textbf{P}- \textbf{B}\textbf{H}\textbf{P} - (\textbf{B}\textbf{H}\textbf{P})^\mathrm{H} \Big) 
	  - \lambda q ,}{}{}
	  \label{eq:lastsd}
\end{mini}
which can be solved by the SD algorithm for a fixed value of $\lambda$.  We can start with $\lambda=1$ and after going through the SD algorithm, we will check the power constraint $\mathrm{tr}(\textbf{P}\textbf{P}^\mathrm{H}) \le q$ to determine if $\lambda$ should be increased or decreased.
By using the bisection method \cite{burden19852}, we can find the best $\lambda$.

We first convert \eqref{eq:lastsd} for a fixed $\lambda$ into a vector optimization problem using  $\textbf{a} = \mathrm{vec}(\textbf{P})$ and $\textbf{h} = \mathrm{vec}\Big( (\textbf{BH})^\mathrm{T}\Big)$. We thereby obtain 
\begin{mini}|l|
	  {\textbf{a}_i\in {\mathcal{P}}^{M \times 1}, i=1,\ldots,K }{\sum_{i=1}^K \Big( \textbf{a}_i^\mathrm{H} \left ( \textbf{H}^\mathrm{H}\textbf{B}^\mathrm{H} \textbf{B}\textbf{H}+\lambda \textbf{I}_M \right )\textbf{a}_i -\textbf{h}_i^\mathrm{T}\textbf{a}_i - \left ( \textbf{h}_i^\mathrm{T}\textbf{a}_i\right )^\mathrm{H}\Big),}{}{}
 \label{eq:modifysd}
\end{mini}
which has $K$ separable objective functions that each only depends on one of the optimization variables.
This feature enables parallel optimization of $\textbf{a}_i$ for $i=1,\ldots,K$, where $\textbf{a}_i$ is a vector with $M$ entries containing elements from the $k$-th column of $\textbf{P}$. Thus, in addition to the more efficient search strategy, the reformulation in \eqref{eq:lastsd} also significantly reduces the dimension of each subproblem. 

By defining $\hat{\textbf{V}} = \textbf{H}^\mathrm{H}\textbf{B}^\mathrm{H} \textbf{B}\textbf{H}+\lambda \textbf{I}_M$, we can obtain the equivalent formulation of each term of the objective function in \eqref{eq:modifysd} as 
\begin{align}
    &\textbf{a}_i^\mathrm{H} \hat{\textbf{V}} \textbf{a}_i -\textbf{h}_i^\mathrm{T}\textbf{a}_i - \left ( \textbf{h}_i^\mathrm{T}\textbf{a}_i\right )^\mathrm{H} \nonumber \\ &= \lVert \textbf{e}_i - \textbf{R}\textbf{a}_i \rVert_2^2 - \textbf{e}_i^\mathrm{H}\textbf{e}_i,
\end{align}
where $\textbf{R} \in \mathbb{C}^{M \times M }$ is obtained from the Cholesky decomposition $\hat{\textbf{V}} = \textbf{R}^\mathrm{H} \textbf{R}$ and $\textbf{e}_i = (\textbf{h}_i^\mathrm{T}\textbf{R}^{-1})^\mathrm{H}$. As $\textbf{e}_i \in \mathbb{C}^{M \times 1 }$ does not depend on the optimization variable, we can rewrite the subproblem with respect to $\textbf{a}_i$ as 
\begin{mini}|l|
	  {\textbf{a}_i\in {\mathcal{P}}^{M \times 1} }{\lVert \textbf{e}_i - \textbf{R}\textbf{a}_i \rVert_2^2.}{}{}
 \label{eq:sdlasteq1}
\end{mini}

The subproblem formulation in \eqref{eq:sdlasteq1} has complex-valued variables, but we have noticed that the run time can be reduced by rewriting it in an equivalent real-valued form called as real-valued SP. Later in the paper, we call real-valued SP as SP scheme. We then need the following definitions:
\begin{align}
 \textbf{e}_{i,\mathbb{R}} & =
\begin{bmatrix}
\mathfrak{R}\{\textbf{e}_{i}^1\}\\
\mathfrak{I}\{\textbf{e}_{i}^1\}\\
\vdots
\\
\mathfrak{R}\{\textbf{e}_{i}^M\}\\
\mathfrak{I}\{\textbf{e}_{i}^M\}
\end{bmatrix}, \quad
 \textbf{a}_{i,\mathbb{R}}=
\begin{bmatrix}
\mathfrak{R}\{\textbf{a}_{i}^1\}\\
\mathfrak{I}\{\textbf{a}_{i}^1\}\\
\vdots
\\
\mathfrak{R}\{\textbf{a}_{i}^M\}\\
\mathfrak{I}\{\textbf{a}_{i}^M\}
 \nonumber
\end{bmatrix}, \quad \text{and } \\
\textbf{R}_\mathbb{R}&=
\begin{bmatrix}
\mathfrak{R}\{\textbf{R}_{1,1}\} & - \mathfrak{I}\{\textbf{R}_{1,1}\}& \ldots\\
\mathfrak{I}\{\textbf{R}_{1,1}\}& \mathfrak{R}\{\textbf{R}_{1,1}\} & \ddots
\\ \vdots & \vdots & \ldots
\end{bmatrix}.
\end{align}
By using these definitions, we can finally rewrite \eqref{eq:sdlasteq1} as \begin{mini}|l|
	  {\textbf{a}_i\in {\mathcal{L}}^{2M \times 1} }{\lVert \textbf{e}_{i,\mathbb{R}} - \textbf{R}_\mathbb{R}\textbf{a}_{i,\mathbb{R}} \rVert_2^2.}{}{}
 \label{eq:sdlasteq2}
\end{mini}
The triangular structure of \eqref{eq:sdlasteq2} allows us to employ the same SD method as in \cite{agrell2002closest}. We call this overall approach \emph{sphere precoding (SP)}, in line with the previous work \cite{jacobsson2017quantized} that adapted the SD method for 1-bit quantized precoding.
The first lattice point explored by the SESD with the infinite radius is always the Babai point \cite{hassibi2005sphere}, which is also referred to as the nulling and canceling (NC) point \cite{foschini1999simplified}. In the viewpoint of the tree search, the branches are examined according to the ascending order of the branch metric in SESD. If one candidate symbol violates the sphere constraint in the Schnorr-Euchner enumeration, SESD can remove the next candidates which do not comply with the constraint condition. Psuedo-code for the SP algorithm is summarized in Algorithm~\ref{alg:SESD}.

\begin{algorithm} 
\SetKwInOut{Input}{Input}
\SetKwInOut{Output}{Output}
\Input{$\textbf{R}_\mathbb{R}$, $\textbf{e}_{i,\mathbb{R}}$, $\mathcal{L}$, $r_{\text{opt}}$, $2M$}
\DontPrintSemicolon

$\textbf{a}_{i,\mathbb{R}} \leftarrow \textbf{0}_{2M \times 1}$, $\hat{\textbf{a}}_{i,\mathbb{R}} \leftarrow \textbf{0}_{2M \times 1}$, $\bar{\textbf{a}}_{i,\mathbb{R}} \leftarrow \textbf{0}_{2M \times 1}$, $m \leftarrow 2M+1$, $\textbf{r}[m] \leftarrow 0$, $\mathrm{state} \leftarrow  \mathrm{down}$\;

\While{$!\Big( (\mathrm{state} == \mathrm{up}) \& \& (m == 2M) \Big)$}{ 
\uIf{$(\mathrm{state} == \mathrm{down})$}{
$m \leftarrow m-1$ \;
\If{$\textbf{R}_\mathbb{R}[m,m] != 0 $}{$\hat{\textbf{a}}_{i,\mathbb{R}}[m] \leftarrow \frac{\textbf{e}_{i,\mathbb{R}}[m]-\textbf{R}_\mathbb{R}[m,m:2M]\bar{\textbf{a}}_{i,\mathbb{R}}[m:2M]}{\textbf{R}_\mathbb{R}[m,m]} $ \tcp{Zero-forcing decision feedback Detector}}

   $\text{Round }\hspace{1mm}\hat{\textbf{a}}_{i,\mathbb{R}}[m] \hspace{1mm} \text{to closest point and set}\hspace{1mm} \bar{\textbf{a}}_{i,\mathbb{R}}[m]\hspace{1mm} \text{to be searched}$\tcp{$\bar{\textbf{a}}_{i,\mathbb{R}}[m] \leftarrow$point} 
    
    $\textbf{s}[m] \leftarrow \Delta \cdot \sgn \big(\hat{\textbf{a}}_{i,\mathbb{R}}[m] - \bar{\textbf{a}}_{i,\mathbb{R}}[m] \big)$ \;
    }
\Else{ $m \leftarrow m+1$ \;
$\bar{\textbf{a}}_{i,\mathbb{R}}[m] \leftarrow \bar{\textbf{a}}_{i,\mathbb{R}}[m] + \textbf{s}[m]$   \; 
$\textbf{s}[m] \leftarrow - \big( \textbf{s}[m] + \Delta \cdot \sgn (\textbf{s}[m]) \big)$ \tcp{Zig-zag implementation} 
\If{ $(\bar{\textbf{a}}_{i,\mathbb{R}}[m] < l_0) || (\bar{\textbf{a}}_{i,\mathbb{R}}[m] >l_{L-1})$}{
$\bar{\textbf{a}}_{i,\mathbb{R}}[m] \leftarrow \bar{\textbf{a}}_{i,\mathbb{R}}[m] + \textbf{s}[m]$   \; 
$\textbf{s}[m] \leftarrow - \big( \textbf{s}[m] + \Delta \cdot \sgn (\textbf{s}[m]) \big)$ \;}
}

 \uIf{$(\bar{\textbf{a}}_{i,\mathbb{R}}[m] < l_0) || (\bar{\textbf{a}}_{i,\mathbb{R}}[m]>l_{L-1})$}{ 
 $\mathrm{state} \leftarrow \mathrm{up}$}
 \Else{
    $\textbf{r}[m] \leftarrow (\textbf{R}_\mathbb{R}[m,m] \times (\bar{\textbf{a}}_{i,\mathbb{R}}[m] - \hat{\textbf{a}}_{i,\mathbb{R}}[m]))^2 + \textbf{r}[m+1] $   \;   
    \uIf{$\textbf{r}[m] < r_{\text{opt}}$}{
    \uIf{(m == 1)} {${\textbf{a}}_{i,\mathbb{R}}[m] \leftarrow \bar{\textbf{a}}_{i,\mathbb{R}}[m]$
    \tcp{Save optimal vector}
    
    $r_{\text{opt}} \leftarrow \textbf{r}[m]$ \tcp{Save new best radius}
    
    $\mathrm{state} \leftarrow \mathrm{up}$}
    \Else{$\mathrm{state} \leftarrow \mathrm{down}$\tcp{Continue to go down in tree}}
    
    }
    \Else{$\mathrm{state} \leftarrow \mathrm{up}$\tcp{Combination violated the sphere constraint}}
}}
\textbf{Return} $\textbf{a}_{i,\mathbb{R}} $
\caption{SESD algorithm for solving \eqref{eq:sdlasteq2}}
\label{alg:SESD}
\end{algorithm}

\begin{remark}
The algorithms developed in this section apply for a fixed matrix $\textbf{B}$ (i.e., fixed $\boldsymbol \beta$). An alternating optimization approach can be taken to also optimize $\textbf{B}$.
At iteration $t = 1$, we can initialize the algorithm with the precoding factor obtained from WF precoding. Specifically,
we use \eqref{eq:betawf} and set $\boldsymbol \beta^1 = \boldsymbol \beta^{WF}$. We can then use our SP algorithm to obtain $\textbf{P}$. Next, we can compute an improved precoding factor $\boldsymbol \beta^{t+1} = v(\textbf{P})$ using \eqref{eq:betaopt}. By repeating this procedure for $t = 2, 3, \ldots$ until convergence (or until a maximum number of iterations as been reached) we can jointly optimize $\textbf{P}$ and $\boldsymbol \beta$. However, our simulations have shown that the performance of the system does not vary a lot between iterations and, hence, we believe that the increased complexity is not worth it.
\end{remark}
\section{Heuristic Quantization-Aware Precoding}\label{sec4sumrate}

Although the SP algorithm is relatively efficient, compared to a general-purpose solver, there is a limit to how large setups ($M$ and $K$) it can handle in real-time applications. Hence, we believe it should primarily be seen as a benchmark for designing lower-complexity precoding schemes. 
In this section, we will develop a heuristic quantization-aware precoding scheme.
The proposed scheme is initiated from the quantization-unaware WF precoding scheme described in Section \ref{sec2unaware}. The precoding matrix elements are then refined in a sequential manner, by exploring alternative ways of quantizing each element.

\begin{figure}[t!]
  \centering
   \begin{overpic}[scale=0.55,unit=1mm]{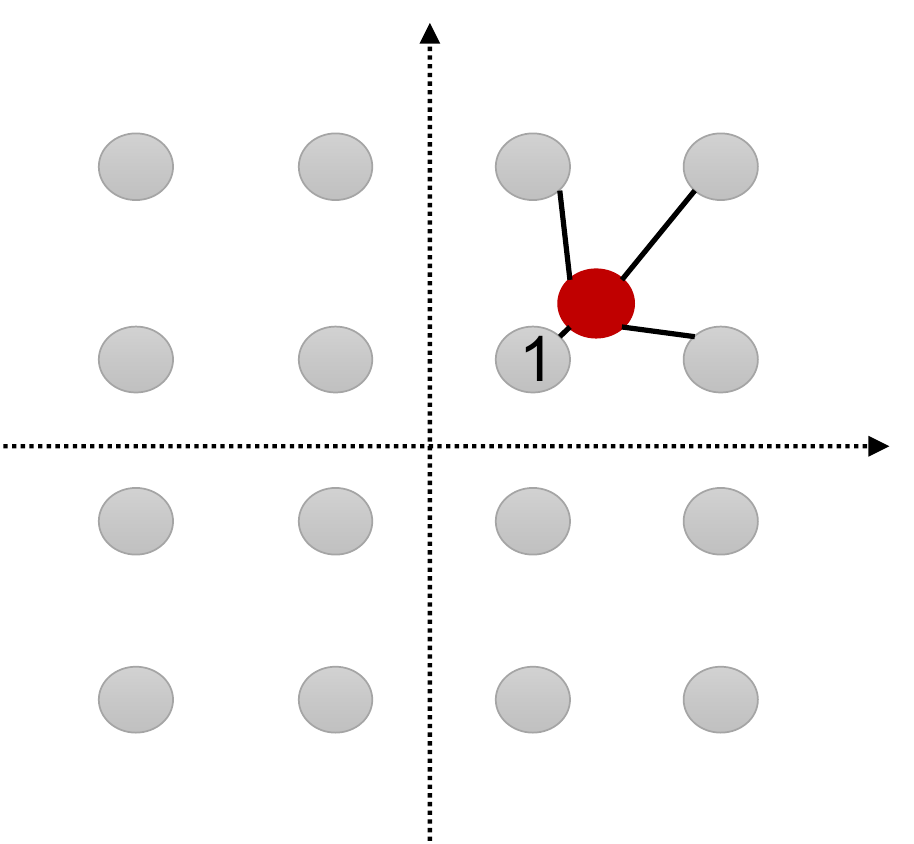}
  \put(100,42){Real part}%
  \put(20,93){Imaginary part}%
   \end{overpic}
\caption{This figure shows the 16 grid points when quantizing using 2 bits per real dimension. The red point indicates an element from a precoding matrix computed using infinite resolution. The quantization-unaware scheme will quantize this element to the closest point, indicated by a ``1''. We propose a heuristic scheme that also considers the second closest quantization label in both the real and imaginary dimensions, and selects the one that maximizes the sum rate.}
\label{fig:heuristic}
\end{figure}

We will first provide an example to motivate the algorithmic design.
Consider the case where 2-bit quantization is used over the fronthaul. There will then be 4 quantization levels per real dimension. Since each element of the precoding matrix is complex-valued, it must be quantized to one of the points on a two-dimensional grid with $4 \times 4 = 16$ points.
Fig.~\ref{fig:heuristic} shows this grid and how a red point (representing a precoding matrix element before quantization) would be quantized to the geometrically closest grid point (marked by ``1'').
The quantization-unaware precoding described in Section~\ref{sec2unaware} does precisely this.
However, we could also consider the second closest quantization levels in both the real and imaginary dimensions. This gives us three alternative ways of quantizing this precoding matrix element. These points are indicated with connecting lines in Fig.~\ref{fig:heuristic}.
Since the interference suppression capability of a precoding method is determined by how the signals from all antennas cancel out at the undesired receivers, it might be acceptable to increase the quantization errors (in a per-element MSE sense) if that reduces the total interference. 
Motivated by this fact, we propose to select the one of the four alternatives that maximizes the sum rate.

The proposed method updates the elements of the quantization-unaware precoding matrix $\textbf{P}$ sequentially, thus, we need to order the elements properly.
We propose to start by updating the column of the precoding matrix corresponding to the UE $m$ with the highest \textit{generated interference} $\mathrm{GI}_k = \sum_{{i=1} , i \ne k}^K |[{\textbf{H}} {\hat{\textbf{P}}}]_{i,k}|^2$, where $\hat{\textbf{P}} = \alpha \textbf{P} = \alpha \mathcal{Q} (\textbf{W}^\mathrm{WF})$ since this might improve the performance the most.\footnote{We have noticed experimentally that this leads to the largest improvement in sum rate at high SNR.} Then for that specific user $k$, for each transmit antenna $m \in \{ 1, \ldots, M \}$, we identify the four nearest points in $\mathcal{P}$ to the element $w_{k,m}^\mathrm{WF}$ from the original unquantized precoding matrix $\textbf{W}^\mathrm{WF}$.
We evaluate the sum rate
\begin{equation}  \sum_{k=1}^K 
\log_2 \left(1+\frac{\big|[{\textbf{H}} \hat{\textbf{P}}]_{k,k}\big|^2}{\sum_{i=1, i  \ne k}^{K}\big|[{\textbf{H}} \hat{\textbf{P}}
]_{k,i}\big|^2 +N_0} \right)
\label{eq:sumrate}
\end{equation}
for the four different $\textbf{P}$ options obtained with $p_{k,m} \in \{ \text{four nearest points to } w_{k,m} \hspace{1mm} \text{in} \hspace{1mm} \mathcal{P} \} $ while all other elements are fixed.
We then replace the corresponding element in $\textbf{P}$ with the option that achieved the largest sum rate.
The rest of the UEs are ordered based on decreasing generated interference and the precoding elements are updated in the same way.

If the complexity of refining the precoding for all $K$ UEs is too large, the heuristic algorithm can be terminated when $S$ UEs have been considered. Any number $1 \leq S \leq K$ can be considered. This approach might be of interest in massive MIMO systems, where the precoding matrix has hundreds of elements. We will evaluate the impact of $S$ in the next section.

\section{Numerical results}\label{sec5num}

In this section, we will evaluate the performance of the optimized and heuristic quantization-aware precoding under different conditions. We will use the sum rate as performance metric, instead of the tightly connected sum MSE that was used in the optimization, since it is more common when measuring end-user performance.
We will compare the proposed methods with quantization-unaware precoding and also the finite-alphabet precoding approach from \cite{castaneda2019finite}.
The entries of the channel matrix $\textbf{H}$ are generated as i.i.d. $\mathcal{CN}(0,1)$, to demonstrate that our algorithms are not relying on any particular channel structure.
The average sum rate is calculated as
\begin{equation}  \sum_{k=1}^K 
\mathbb{E} \left[\log_2 \left(1+\frac{\big|[{\textbf{H}} \hat{\textbf{P}}]_{k,k}\big|^2}{\sum_{i=1, i  \ne k}^{K}\big|[{\textbf{H}} \hat{\textbf{P}}
]_{k,i}\big|^2 +N_0} \right)\right]
\label{eq:sumrateaverage}
\end{equation}
using Monte Carlo simulations for the case of perfect CSI at the receiver.
Our baseline simulation setup has $M=16$ BS antennas and $K=4$ UEs. These UEs have a common SNR that we define as $\rho=\frac{q}{N_0}$.
We will compare different precoding schemes as a function of the SNR and the number of quantization levels $L$. We will also compare quantized precoding with the ideal infinite-resolution case (without quantization). 
In some simulations, we will deviate from the baseline scenario by considering varying SNRs and larger values of $M$ and $K$.

\begin{figure}[!t]
        \centering
        \vspace{-30mm}\includegraphics[scale=0.45
        ]{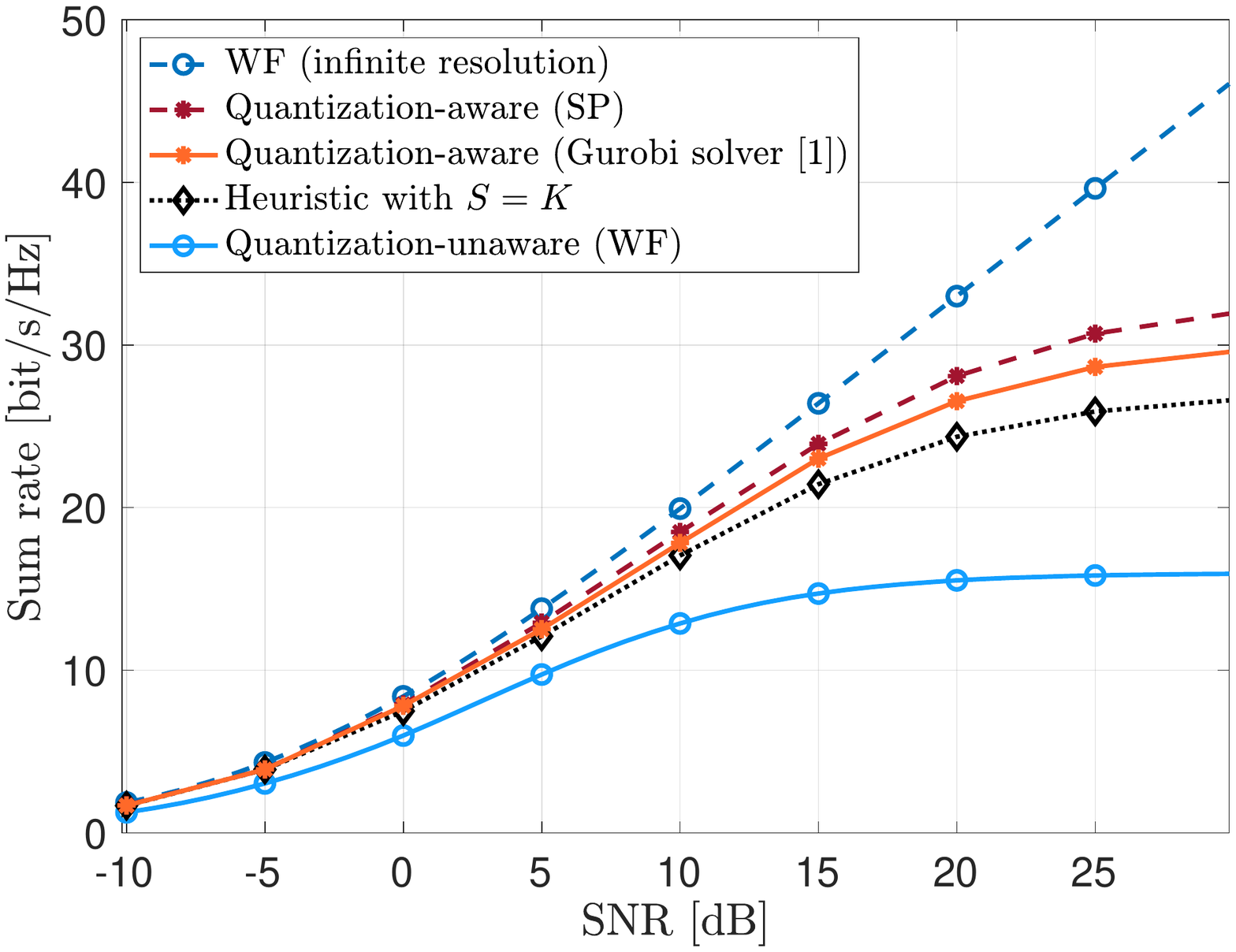}
        \vspace{-30mm}
        \caption{The average sum rate versus the SNR for different precoding schemes. We assume that the AAS has $M=16$ antennas and serves $K=4$ UEs, with $L=8$ quantization levels.}
        \label{fig:sum}
\end{figure} 


Fig.~\ref{fig:sum} shows the average sum rate as a function of the SNR for the following schemes:
\begin{enumerate}
\item WF (infinite resolution), which uses WF precoding without quantization;
\item Quantization-aware (SP), which uses the algorithm proposed in Section~\ref{subsec:SP};
\item Quantization-aware (Preliminary work \cite{yasaman2022}), which considers the less refined problem formulation from the conference version of this paper;
\item Heuristic, which is the proposed heuristic quantization-aware algorithm from Section~\ref{sec4sumrate};
\item Quantization-unaware (WF), which is WF precoding followed by quantization.
\end{enumerate}
The number of quantization levels is $L=8$. The infinite-resolution WF precoding outperforms all the quantized precoding schemes, and the gap increases linearly (in dB scale) at high SNR since the quantization effect imposes a fundamental limit on the interference suppression capability.
Nevertheless, the proposed quantization-aware precoding performs remarkably better than the quantization-unaware WF precoding, with almost a doubling of the sum rate.
The proposed SP approach also outperforms the quantization-aware algorithm from our preliminary work \cite{yasaman2022}, which assigned the same precoding factor to all UEs and used CVX/Gurobi to solve the MSE minimization.
It is worth mentioning that the proposed SP approach only finds a local optimum, due to the sequential search over the Lagrange multiplier, but we noticed that this only reduces the sum rate by around five percent.
The proposed heuristic algorithm from Section~\ref{sec4sumrate} with $S=K$ is not reaching the same performance as the SP approach, but  still performs vastly better than quantization-unaware WF precoding.

\subsection{Computational Complexity}

We will now compare the computational complexity orders of the considered algorithms. We measure the complexity order in terms of the number of real-valued multiplications, using big-$O$ notation. The original problem in \eqref{eq:mmse} is combinatorial and can be solved by an exhaustive search with a complexity of $O(L^{2MK})$.
The mixed-integer convex reformulation in \eqref{eq:realinteger} finds a suboptimal solution for a fixed $\boldsymbol{\beta}$ and can be solved by CVX using the Gurobi solver. While the exact complexity is substantially lower than with an exhaustive search, the complexity scaling is the same.
We propose an SP-based implementation that has substantially reduced complexity since it enables parallel optimization of $K$ variables in \eqref{eq:sdlasteq2}. Since the implementation relies on similar steps as in classical SD, the algorithm's average complexity is at the order of $O(KL^{2\gamma M})$ for some $0 \le \gamma \le 1$ \cite{jalden2005complexity}. 
Finally, we notice that the computational complexity of both the quantization-unaware precoding and optimized heuristic approach (presented in this section) scales as $O(MK^2 + K^3)$ since the most complex operations are matrix multiplications and divisions. This is the same complexity order as with infinite WF precoding, even if the exact complexity is slightly higher. 

Table~\ref{table:time} compares the average run time of the different quantization-aware precoding schemes. We used a 13-inch Macbook with an M1 processor and 8 GB memory. The numbers are averaged over 100 random channel realizations. Using the general-purpose CVX/Gurobi solver results in the largest run time, while the SP formulation is $3 \cdot 10^4$ times faster. These incredible differences demonstrate the importance of exploiting the problem structure when crafting the optimization algorithm.
Finally, we notice that the heuristic algorithm is $300$ times faster than the proposed SP approach and reaches sub-second
numbers, thus making it appropriate for real-time applications.

\begin{table}[!t]
\centering
\caption{Average run time of different quantization-aware precoding schemes.}
\begin{tabular}{|ll|}
\hline 
\multicolumn{1}{|l|}{Algorithm}  & Total run time (sec) \\ \hline
\multicolumn{1}{|l|}{CVX/Gurobi} &    86905          \\ \hline
\multicolumn{1}{|l|}{SP}    &   3.09               \\ \hline
\multicolumn{1}{|l|}{Heuristic with $S=K$}    &   0.0109               \\ \hline
\end{tabular} \label{table:time}
\end{table}

\subsection{Finite-alphabet WF precoding}

\begin{figure}[!t]
        \centering
        \vspace{-35mm}\includegraphics[scale=0.45]{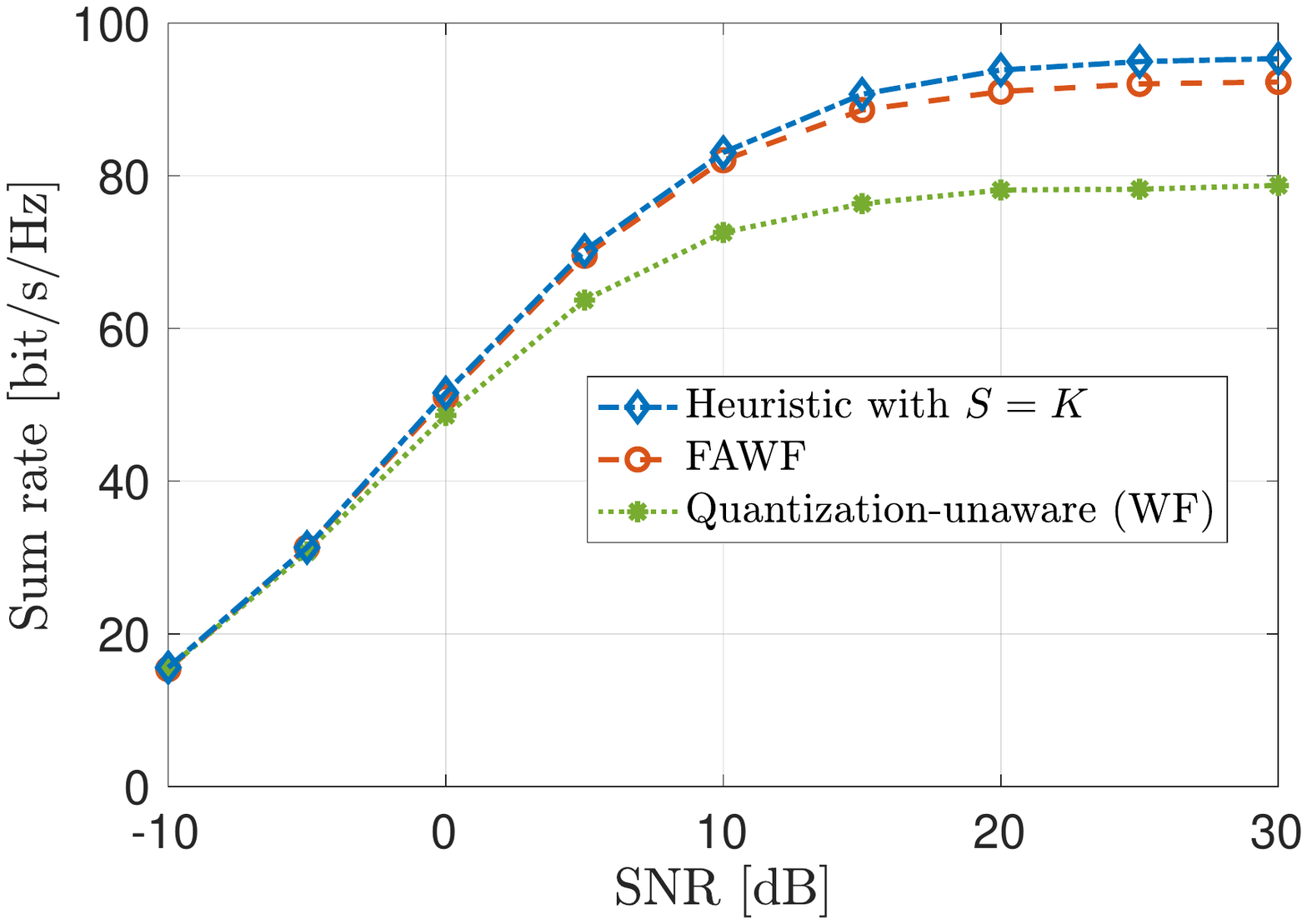}
        \vspace{-30mm}
        \caption{The average sum rate versus the SNR. We compare the FAWF algorithm from \cite{castaneda2019finite}, our proposed heuristic quantization-aware method and the baseline quantization-unaware WF precoding. The setup contains $M=256$, $K=16$, and $L=2$.}
        \label{fig:fawf}
\end{figure}

In \cite{castaneda2019finite}, the authors propose a
quantization-aware precoding scheme called finite-alphabet WF precoding (FAWP). They propose to use a precoding matrix on the form 
\begin{equation}
    \textbf{P}^{\text{FAWP}} = \textbf{Q}\textbf{B}^{q},
    \label{eq:fawp}
\end{equation}
where $\textbf{Q} \in {\mathcal{P}}^{M \times K}$ is a low-resolution matrix with entries taken
from the finite alphabet $\mathcal{P}$ and $\textbf{B}^q$ is a ${K\times K}$ diagonal matrix with per-UE scaling factors. They call $\textbf{P}^{\text{FAWP}}$ the pre-FAWP matrix. 
They use a forward-backward splitting algorithm to approximately solve an MMSE problem for precoding matrices of the form \eqref{eq:fawp} and show numerically that the result outperforms the quantization-unaware WF precoding described in Section \ref{sec2unaware}. We have noticed that this algorithm requires careful parameter tuning and \cite{castaneda2019finite} uses machine learning for that purpose. 
In Fig.~\ref{fig:fawf}, the pre-FAWF-FBS algorithm from \cite{castaneda2019finite} is compared with our heuristic algorithm from Section~\ref{sec4sumrate}. We consider a 1-bit alphabet with $M=256$ antennas and $K=16$ UEs as in \cite{castaneda2019finite}, to reuse the neural-network-based tuning from that paper. This can be viewed as a massive MIMO setup.
As expected, the sum rate with the two quantization-aware schemes converge to upper limits at high SNR, but the limits are substantially higher than with quantization-unaware WF.
We notice that our new heuristic scheme provides slightly higher rates than FAWF, a slight reduction in complexity and also alleviates the need for parameter tuning (e.g., based on machine learning).

\begin{figure}[!t]
        \centering
        \vspace{-35mm}\includegraphics[scale=0.45]{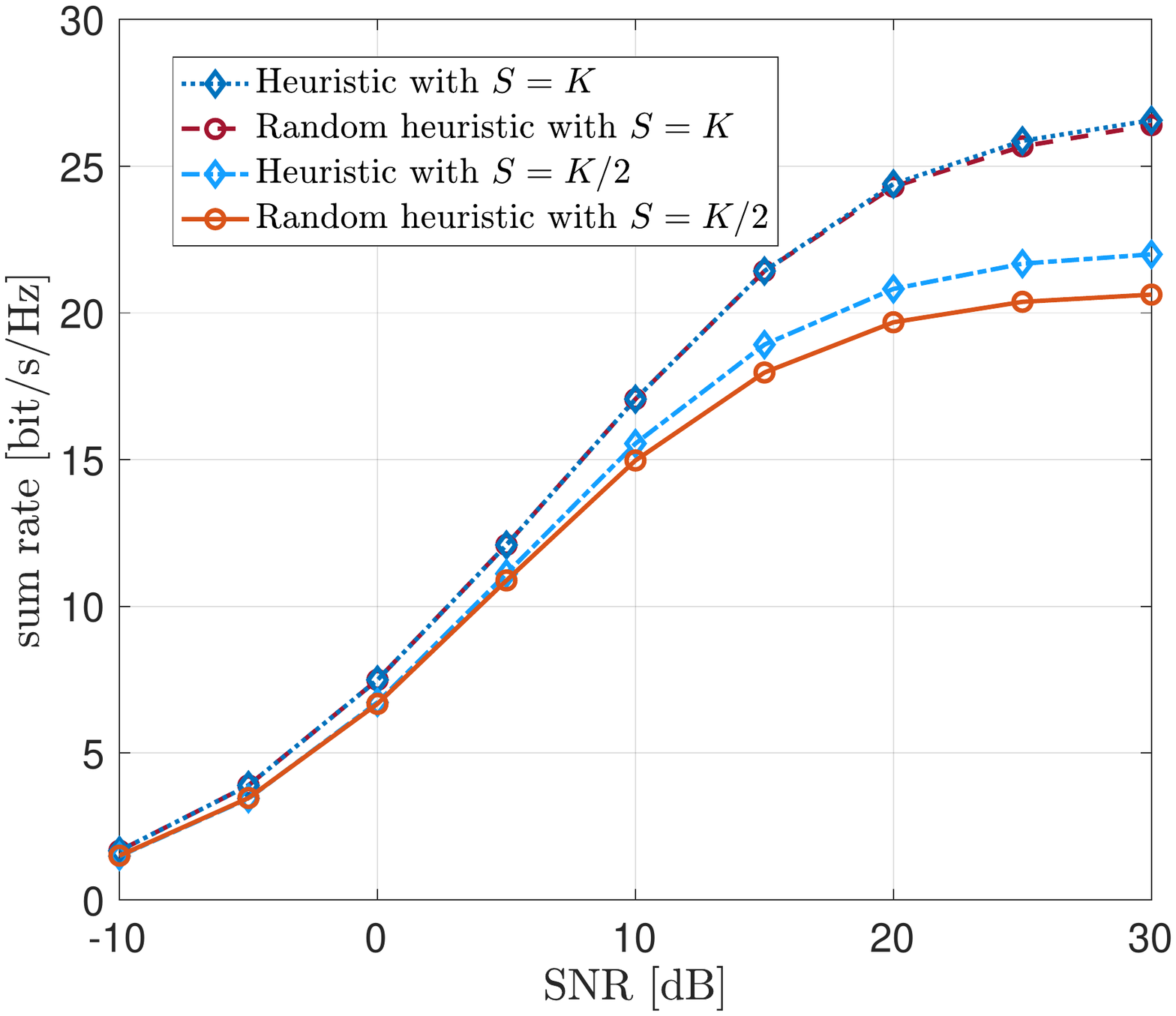}
        \vspace{-30mm}
        \caption{The average sum rate versus the SNR for proposed heuristic scheme with different UE ordering. The setup is the same as in Fig. \ref{fig:sum}.}
      \label{fig:heu}
\end{figure}

Fig.~\ref{fig:heu} evaluates different variations of our heuristic quantization-aware precoding scheme. The setup contains $M=16$ BS antennas and $K=4$ UEs with $L=8$ quantization levels. We show the average sum rate versus the SNR for $S=K$ and $S=K/2$; that is, we refine the precoding vectors for all UEs or half of them. We also compare the proposed UE ordering based on the largest generated interference with a random UE ordering. We first notice that the UE ordering has negligible impact if $S=K$, since the proposed ordering is no better than  random ordering. However, if we reduce $S$ to lower the computational complexity, which is relevant for massive MIMO scenarios, then the UE ordering becomes important. In the case of $S=K/2$, the proposed ordering that only refines the precoding vectors for the UEs that generates the most interference, performs significantly better at high SNR than the random ordering.  
When developing these results, we also considered an alternative ordering of the UEs based on which ones receive the largest amount of interference.
This alternative leads to slightly better rates at low SNR, since UEs with bad channel conditions are then benefiting more from increasing their own beamforming gains than reducing the interference that others cause to them. However, the difference is so small that we didn't include these curves in the figure.

\subsection{Significance of degrees-of-freedom analysis}

\begin{figure}[!t]
        \centering
        \vspace{-35mm}\includegraphics[scale=0.45]{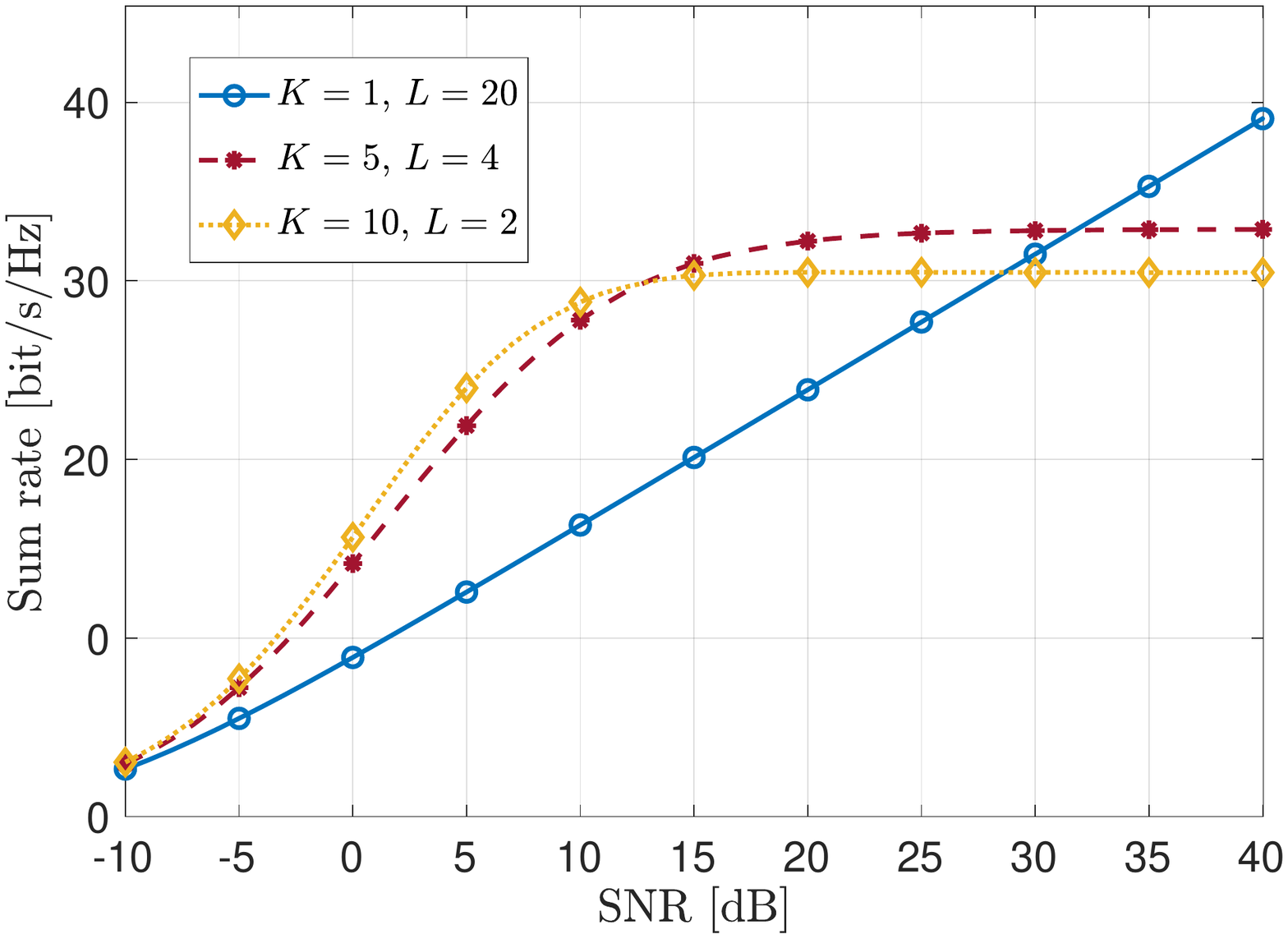}
      \vspace{-35mm}
        \caption{The average sum rate versus the SNR for a fixed number of BS antennas $M=16$. We are varying $K$ and $L$ such that $K \cdot L = 20$ to show how the sum rate is maximized by different numbers of UEs at different SNRs.}
        \label{fig:fix}
\end{figure}

The finite-resolution quantization makes it generally impossible to fully remove interference at high SNR.
This implies that the spatial degrees-of-freedom of the considered MU-MIMO system collapses to $0$ whenever we try to serve multiple UEs, similar to the works such as \cite{jindal2006mimo} on feedback quantization.
However, one can always achieve a spatial degrees-of-freedom of $1$ by only serving one UE at a time. This type of scheduling must be optimal from a sum rate perspective when the SNR becomes sufficiently large.
Moreover, since the fronthaul signaling is determined by the total number of bits needed to represent the precoding matrix, one could potentially increase the resolution per element when reducing the number of UEs.
To study the practical significance of these theoretical insights, Fig.~\ref{fig:fix} presents the average sum rate with the 
proposed quantization-aware SP as a function of the SNR for $M=16$ BS antennas and different combinations of $K$ and $L$, such that $K \cdot L = 20$. This number defines the fronthaul signaling load.
At low and medium SNR, the sum rate is maximized by serving many UEs. At high SNR, serving a lower number of UEs with a higher-resolution quantizer outperforms the opposite case. This is because the system is heavily interference-limited, which can be partially resolved by simultaneously increasing $L$ and decreasing $K$.
For each SNR value, there is an optimal number of UEs that maximizes the sum rate and it is therefore imperative to schedule the right number of UEs.
Despite the fact that single-user transmission prevails at high SNR, MU-MIMO remains the preferable case in practice since the crossing point is at about 30 bit/s/Hz, which would require enormous constellation sizes for a single UE. Practical transceiver impairments typically become the limiting factor if one tries to operate beyond 10-12 bit/s/Hz (i.e., 1024-QAM to 4096-QAM).

\subsection{Imperfect CSI}

\begin{figure}[!t]
        \centering
        \vspace{-5mm}\includegraphics[scale=0.2]{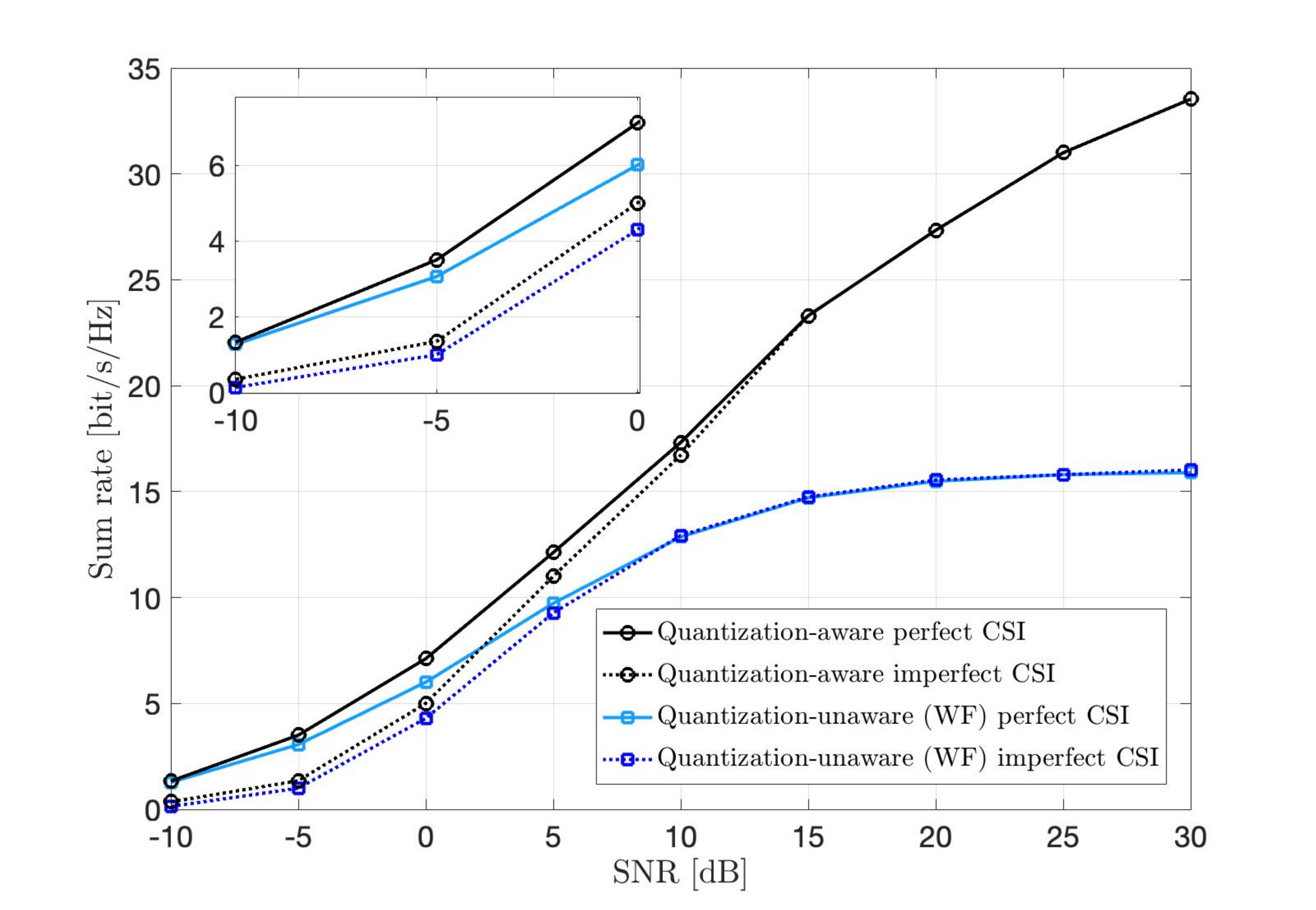}
       
        \caption{The average sum rate versus the  SNR with either perfect or imperfect CSI at the BBU. We consider the same setup as in Fig. \ref{fig:sum}.}
        \label{fig:imperfect}
\end{figure}

Fig.~\ref{fig:imperfect} considers the average sum rate versus the SNR for either perfect CSI at the BBU (as assumed previously in this section) or imperfect CSI calculated based on \eqref{eq:sumrateaverage}. The BBU applies the same algorithms as if the CSI was perfect.
The imperfect CSI results in a sum rate reduction at low SNR, but the proposed quantization-aware precoding  still achieves higher sum rate than the quantization-unaware alternatives.
At high SNR, the performance gap between the perfect and imperfect CSI cases vanishes since the CSI quality also improves with the SNR.
This figure demonstrates that although we have considered perfect CSI when developing our algorithms, it is straightforward to apply the same methodology in situations with imperfect CSI. The CSI imperfections will only be a limiting factor at low SNR, while the finite-resolution quantization is the main limiting factor at high SNR.

\subsection{Different path losses}

All the simulation results that we have presented until now consider that all the UEs have the same SNR $\rho=\frac{q}{N_0}$, which means that they have the same path loss. In real scenarios, each UE will have a different path loss depending on its physical location.
We will now evaluate if the proposed precoding algorithms work well also in that case by revisiting the setup from Fig.~\ref{fig:sum} but with varying SNRs.
We assume the SNR values are equally spaced (in dB) within a certain SNR interval, which is 10 dB in our case. We can thereby change the median SNR value among the UEs, while keeping the SNR variations around this value fixed.
This means that for the $K$ UEs, we divide the 10 dB interval by the number of UEs, $K$, and then assign the UEs with the SNR values $ (\text{median SNR} + k \frac{10}{K})$ dB  for $k \in \{ -\frac{K}{2}, -\frac{K}{2}+1, \ldots, \frac{K}{2}  \}$.

\begin{figure}[!t]
        \centering
        \vspace{-35mm}\includegraphics[scale=0.45]{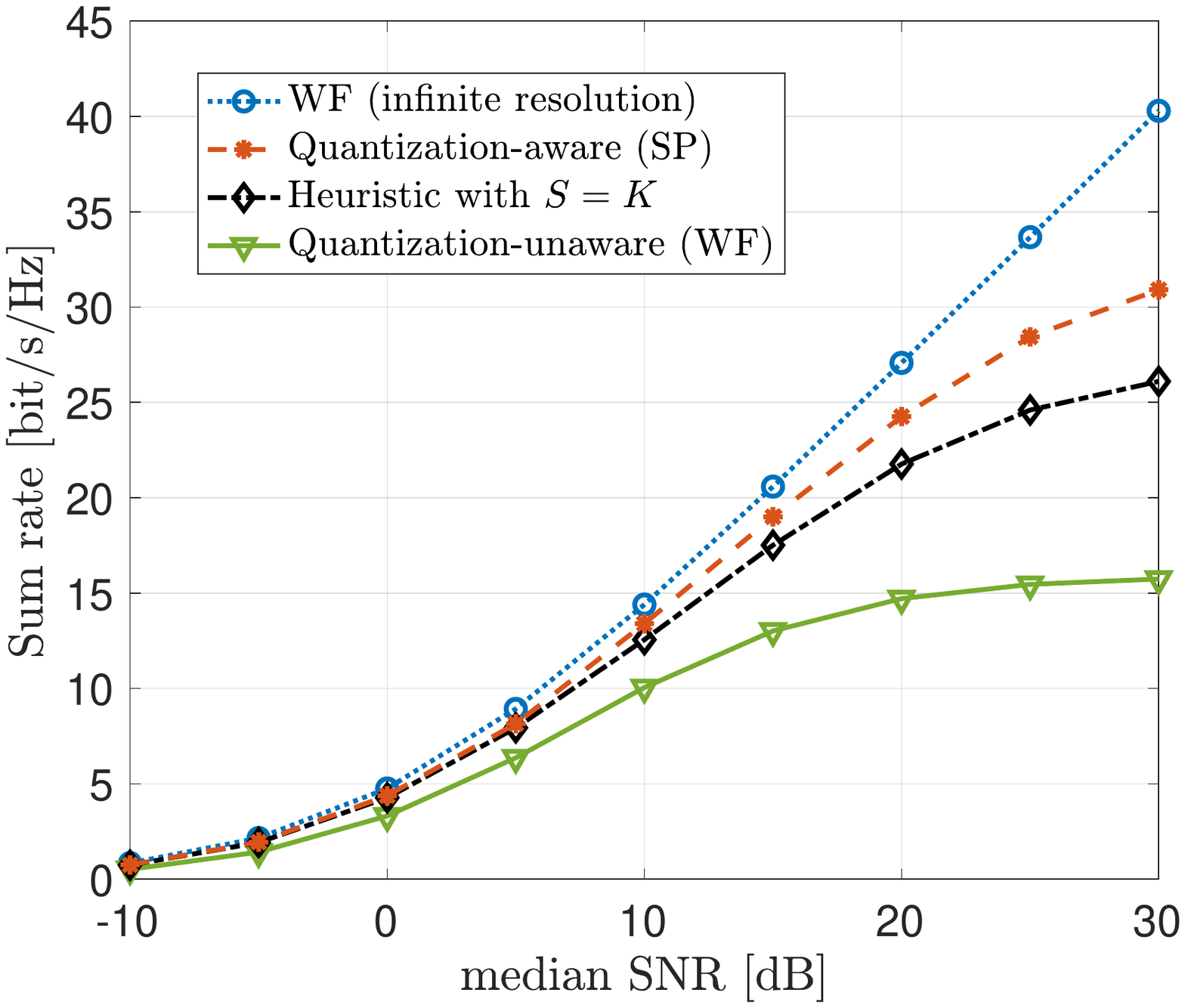}
        \vspace{-30mm}
        \caption{The average sum rate versus the median SNR when there are different path losses among the UEs. We consider the same setup as in Fig. \ref{fig:sum}.}
        \label{fig:pathloss}
\end{figure}

Fig.~\ref{fig:pathloss} presents the average sum rate versus the median SNR among the UEs in the revised setup. 
By comparing these sum rates with Fig.~\ref{fig:sum}, we can notice two things. Firstly, the sum rate reduces when introducing SNR variations around the median since the UEs with reduced SNRs will lose more in rate than the UEs with increased SNRs will benefit. This is a natural consequence of the concave logarithm in the rate expression.
Secondly, the qualitative difference between the different curves remain unchanged. Hence, the proposed quantization-aware precoding methods work well also in scenarios with SNR variations among the UEs.

\section{Conclusions} \label{sec6con}

5G sites often consist of an AAS connected to a BBU via a digital fronthaul with limited capacity. In the downlink, the finite-constellation data symbols can be sent to the AAS without quantization, but the precoding matrix that is computed at the BBU must be quantized to finite precision.
We have introduced a novel framework for quantization-aware precoding, where the BBU uses the quantizer structure to select the best finite-precision MU-MIMO precoding matrix that require no further quantization. In particular, we formulated the MSE-minimizing precoding problem and developed an efficient optimization algorithm to solve it based on the SD methodology. The run time is reduced by four orders-of-magnitude compared to a direct CVX/Gurobi implementation.
We also proposed a heuristic quantization-aware precoding scheme that provides competitive sum rates with even lower computational complexity.
We have shown numerically that the proposed quantization-aware precoding schemes outperform the baseline 
quantization-unaware precoding, where the optimal precoding for the infinite-resolution case is selected and then quantized. The improved interference mitigation capability provided by quantization-awareness gives a large sum rate gain at medium and large SNRs, despite the fact that the maximum spatial degrees-of-freedom is limited to one. While the precoding framework was developed under a perfect CSI assumption, we also showed that it can be applied under imperfect CSI and that the CSI errors only have a significant impact over the quantization effect at low SNR.

\bibliographystyle{IEEEtran}
\bibliography{IEEEabrv,refrences}

\begin{thebibliography}{10}
\providecommand{\url}[1]{#1}
\csname url@samestyle\endcsname
\providecommand{\newblock}{\relax}
\providecommand{\bibinfo}[2]{#2}
\providecommand{\BIBentrySTDinterwordspacing}{\spaceskip=0pt\relax}
\providecommand{\BIBentryALTinterwordstretchfactor}{4}
\providecommand{\BIBentryALTinterwordspacing}{\spaceskip=\fontdimen2\font plus
\BIBentryALTinterwordstretchfactor\fontdimen3\font minus
  \fontdimen4\font\relax}
\providecommand{\BIBforeignlanguage}[2]{{%
\expandafter\ifx\csname l@#1\endcsname\relax
\typeout{** WARNING: IEEEtran.bst: No hyphenation pattern has been}%
\typeout{** loaded for the language `#1'. Using the pattern for}%
\typeout{** the default language instead.}%
\else
\language=\csname l@#1\endcsname
\fi
#2}}
\providecommand{\BIBdecl}{\relax}
\BIBdecl

\bibitem{yasaman2022}
Y.~Khorsandmanesh, E.~Bj\"ornson, and J.~Jald{\'e}n, ``Quantization-aware
  precoding for {MU-MIMO} with limited-capacity fronthaul,'' in \emph{IEEE Int.
  Conf. Acoust., Speech, Signal Process. (ICASSP)}, 2022.

\bibitem{Swales1990a}
S.~C. Swales, M.~A. Beach, D.~J. Edwards, and J.~P. McGeehan, ``The performance
  enhancement of multibeam adaptive base-station antennas for cellular land
  mobile radio systems,'' \emph{{IEEE} Trans. Veh. Technol.}, vol.~39, no.~1,
  pp. 56--67, Feb. 1990.

\bibitem{Gesbert2007a}
D.~Gesbert, M.~Kountouris, R.~W. Heath, C.-B. Chae, and T.~S\"alzer, ``Shifting
  the {MIMO} paradigm,'' \emph{{IEEE} Signal Process. Mag.}, vol.~24, no.~5,
  pp. 36--46, Sep. 2007.

\bibitem{bjornson2017massive}
E.~Bj{\"o}rnson, J.~Hoydis, and L.~Sanguinetti, ``Massive {MIMO} networks:
  Spectral, energy, and hardware efficiency,'' \emph{Foundations and Trends in
  Signal Processing}, vol.~11, no. 3-4, pp. 154--655, 2017.

\bibitem{asplund2020advanced}
H.~Asplund, D.~Astely, P.~von Butovitsch, T.~Chapman, M.~Frenne,
  F.~Ghasemzadeh, M.~Hagstr{\"o}m, B.~Hogan, G.~J{\"o}ngren, J.~Karlsson
  \emph{et~al.}, \emph{Advanced Antenna Systems for {5G} Network Deployments:
  Bridging the Gap Between Theory and Practice}.\hskip 1em plus 0.5em minus
  0.4em\relax Academic Press, 2020.

\bibitem{Bjornson2019d}
E.~Bj\"ornson, L.~Sanguinetti, H.~Wymeersch, J.~Hoydis, and T.~L. Marzetta,
  ``Massive {MIMO} is a reality---{W}hat is next? {F}ive promising research
  directions for antenna arrays,'' \emph{Digital Signal Processing}, vol.~94,
  pp. 3--20, Nov. 2019.

\bibitem{peng2015fronthaul}
M.~Peng, C.~Wang, V.~Lau, and H.~V. Poor, ``Fronthaul-constrained cloud radio
  access networks: Insights and challenges,'' \emph{IEEE Wireless
  Communications}, vol.~22, no.~2, pp. 152--160, 2015.

\bibitem{wenk2010mimo}
M.~Wenk, \emph{{MIMO-OFDM} testbed: Challenges, implementations, and
  measurement results}, ser. Series in microelectronics.\hskip 1em plus 0.5em
  minus 0.4em\relax Hartung-Gorre, 2010.

\bibitem{Zhang2012a}
W.~Zhang, ``A general framework for transmission with transceiver distortion
  and some applications,'' \emph{{IEEE} Trans. Commun.}, vol.~60, no.~2, pp.
  384--399, Feb. 2012.

\bibitem{Bjornson2014a}
E.~Bj{\"{o}}rnson, J.~Hoydis, M.~Kountouris, and M.~Debbah, ``Massive {MIMO}
  systems with non-ideal hardware: Energy efficiency, estimation, and capacity
  limits,'' \emph{{IEEE} Trans. Inf. Theory}, vol.~60, no.~11, pp. 7112--7139,
  Nov. 2014.

\bibitem{Mollen2018a}
C.~Moll\'{e}n, U.~Gustavsson, T.~Eriksson, and E.~G. Larsson, ``Impact of
  spatial filtering on distortion from low-noise amplifiers in massive {MIMO}
  base stations,'' \emph{{IEEE} Trans. Commun.}, vol.~66, no.~12, pp.
  6050--6067, Dec. 2018.

\bibitem{aghdam2020distortion}
S.~R. Aghdam, S.~Jacobsson, U.~Gustavsson, G.~Durisi, C.~Studer, and
  T.~Eriksson, ``Distortion-aware linear precoding for massive {MIMO} downlink
  systems with nonlinear power amplifiers,'' \emph{unpublished paper},
  [Online]. Available: https://arxiv.org/pdf/2012.13337.pdf, 2020.

\bibitem{mollen2016uplink}
C.~Moll\'{e}n, J.~Choi, E.~G. Larsson, and R.~W. Heath, ``Uplink performance of
  wideband massive {MIMO} with one-bit {ADCs},'' \emph{IEEE Trans. Wirel.
  Commun.}, vol.~16, no.~1, pp. 87--100, Jan. 2016.

\bibitem{studer2016quantized}
C.~Studer and G.~Durisi, ``Quantized massive {MU-MIMO-OFDM} uplink,''
  \emph{{IEEE} Trans. Commun.}, vol.~64, no.~6, pp. 2387--2399, Jun. 2016.

\bibitem{jacobsson2017quantized}
S.~Jacobsson, G.~Durisi, M.~Coldrey, T.~Goldstein, and C.~Studer, ``Quantized
  precoding for massive {MU-MIMO},'' \emph{{IEEE} Trans. Commun.}, vol.~65,
  no.~11, pp. 4670--4684, Nov. 2017.

\bibitem{mezghani2009transmit}
A.~Mezghani, R.~Ghiat, and J.~A. Nossek, ``Transmit processing with low
  resolution {D/A}-converters,'' in \emph{2009 16th IEEE International
  Conference on Electronics, Circuits and Systems-(ICECS 2009)}.\hskip 1em plus
  0.5em minus 0.4em\relax IEEE, 2009, pp. 683--686.

\bibitem{jacobsson2019linear}
S.~Jacobsson, G.~Durisi, M.~Coldrey, and C.~Studer, ``Linear precoding with
  low-resolution {DACs} for massive {MU-MIMO-OFDM} downlink,'' \emph{{IEEE}
  Trans. Wireless Commun.}, vol.~18, no.~3, pp. 1595--1609, Mar. 2019.

\bibitem{castaneda2019finite}
O.~Casta{\~n}eda, S.~Jacobsson, G.~Durisi, T.~Goldstein, and C.~Studer,
  ``Finite-alphabet {Wiener} filter precoding for mmwave massive {MU-MIMO}
  systems,'' in \emph{2019 53rd Asilomar Conference on Signals, Systems, and
  Computers}.\hskip 1em plus 0.5em minus 0.4em\relax IEEE, 2019, pp. 178--183.

\bibitem{tabeshnezhad2021reduced}
A.~Tabeshnezhad, A.~L. Swindlehurst, and T.~Svensson, ``Reduced complexity
  precoding for one-bit signaling,'' \emph{IEEE Transactions on Vehicular
  Technology}, vol.~70, no.~2, pp. 1967--1971, 2021.

\bibitem{masouros2010correlation}
C.~Masouros, ``Correlation rotation linear precoding for {MIMO} broadcast
  communications,'' \emph{IEEE Transactions on Signal Processing}, vol.~59,
  no.~1, pp. 252--262, 2010.

\bibitem{tsinos2018symbol}
C.~G. Tsinos, A.~Kalantari, S.~Chatzinotas, and B.~Ottersten, ``Symbol-level
  precoding with low resolution dacs for large-scale array {MU-MIMO} systems,''
  in \emph{2018 IEEE 19th International Workshop on Signal Processing Advances
  in Wireless Communications (SPAWC)}.\hskip 1em plus 0.5em minus 0.4em\relax
  IEEE, 2018, pp. 1--5.

\bibitem{Love2008a}
D.~Love, R.~Heath, V.~Lau, D.~Gesbert, B.~Rao, and M.~Andrews, ``An overview of
  limited feedback in wireless communication systems,'' \emph{{IEEE} J. Sel.
  Areas Commun.}, vol.~26, no.~8, pp. 1341--1365, Oct. 2008.

\bibitem{jindal2006mimo}
N.~Jindal, ``{MIMO} broadcast channels with finite-rate feedback,''
  \emph{{IEEE} Trans. Inf. Theory}, vol.~52, no.~11, pp. 5045--5060, Oct. 2006.

\bibitem{park2013joint}
S.-H. Park, O.~Simeone, O.~Sahin, and S.~Shamai, ``Joint precoding and
  multivariate backhaul compression for the downlink of cloud radio access
  networks,'' \emph{{IEEE} Trans. Signal Process.}, vol.~61, no.~22, pp.
  5646--5658, Nov. 2013.

\bibitem{simeone2009downlink}
O.~Simeone, O.~Somekh, H.~V. Poor, and S.~Shamai, ``Downlink multicell
  processing with limited-backhaul capacity,'' \emph{EURASIP J. Adv. Singal
  Process.}, vol. 2009, pp. 1--10, Jun. 2009.

\bibitem{park2014fronthaul}
S.-H. Park, O.~Simeone, O.~Sahin, and S.~S. Shitz, ``Fronthaul compression for
  cloud radio access networks: Signal processing advances inspired by network
  information theory,'' \emph{{IEEE} Signal Process. Mag.}, vol.~31, no.~6, pp.
  69--79, Nov. 2014.

\bibitem{parida2018downlink}
P.~Parida, H.~S. Dhillon, and A.~F. Molisch, ``Downlink performance analysis of
  cell-free massive {MIMO} with finite fronthaul capacity,'' in \emph{88th
  Vehicular Technology Conference (VTC-Fall)}, Chicago, IL, USA, 2018, pp.
  1--6.

\bibitem{huang2018functional}
Y.~Huang, C.~Lu, M.~Berg, and P.~{\"O}dling, ``Functional split of zero-forcing
  based massive {MIMO} for fronthaul load reduction,'' \emph{IEEE Access},
  vol.~6, pp. 6350--6359, 2018.

\bibitem{park2017massive}
S.~Park, H.~Lee, C.-B. Chae, and S.~Bahk, ``Massive {MIMO} operation in
  partially centralized cloud radio access networks,'' \emph{Computer
  Networks}, vol. 115, pp. 54--64, 2017.

\bibitem{wolsey2007mixed}
L.~A. Wolsey, ``Mixed integer programming,'' \emph{Wiley Encyclopedia of
  Computer Science and Engineering}, pp. 1--10, Mar. 2007.

\bibitem{grant2014cvx}
M.~Grant and S.~Boyd, ``Cvx: Matlab software for disciplined convex
  programming, version 2.1,'' 2014.

\bibitem{CoeyLubinVielma2018}
C.~Coey, M.~Lubin, and J.~P. Vielma, ``Outer approximation with conic
  certificates for mixed-integer convex problems,'' \emph{arXiv preprint
  arXiv:1808.05290}, 2018.

\bibitem{gurobi2020reference}
G.~O. Gurobi, ``Reference manual, {Gurobi} optimization,'' 2020.

\bibitem{hui2001unifquantized}
D.~Hui and D.~L. Neuhoff, ``Asymptotic analysis of optimal fixed-rate uniform
  scalar quantization,'' \emph{{IEEE} Trans. Inf. Theory}, vol.~47, no.~3, pp.
  957--977, Mar. 2001.

\bibitem{joham2005linear}
M.~Joham, W.~Utschick, and J.~A. Nossek, ``Linear transmit processing in {MIMO}
  communications systems,'' \emph{{IEEE} Trans. Signal Process.}, vol.~53,
  no.~8, pp. 2700--2712, Aug. 2005.

\bibitem{bjornson2014beamforming}
E.~Bj\"ornson, M.~Bengtsson, and B.~Ottersten, ``Optimal multiuser transmit
  beamforming: A difficult problem with a simple solution structure [lecture
  notes],'' \emph{IEEE Signal Processing Magazine}, vol.~31, no.~4, pp.
  142--148, 2014.

\bibitem{hassibi2005sphere}
B.~Hassibi and H.~Vikalo, ``On the sphere-decoding algorithm {I}. expected
  complexity,'' \emph{IEEE transactions on signal processing}, vol.~53, no.~8,
  pp. 2806--2818, 2005.

\bibitem{murugan2006unified}
A.~D. Murugan, H.~El~Gamal, M.~O. Damen, and G.~Caire, ``A unified framework
  for tree search decoding: Rediscovering the sequential decoder,'' \emph{IEEE
  Transactions on Information Theory}, vol.~52, no.~3, pp. 933--953, 2006.

\bibitem{jalden2005complexity}
J.~Jald{\'e}n and B.~Ottersten, ``On the complexity of sphere decoding in
  digital communications,'' \emph{IEEE transactions on signal processing},
  vol.~53, no.~4, pp. 1474--1484, 2005.

\bibitem{agrell2002closest}
E.~Agrell, T.~Eriksson, A.~Vardy, and K.~Zeger, ``Closest point search in
  lattices,'' \emph{IEEE transactions on information theory}, vol.~48, no.~8,
  pp. 2201--2214, 2002.

\bibitem{burden19852}
R.~L. Burden and J.~D. Faires, ``2.1 the bisection algorithm,'' \emph{Numerical
  analysis}, pp. 46--52, 1985.

\bibitem{foschini1999simplified}
G.~J. Foschini, G.~D. Golden, R.~A. Valenzuela, and P.~W. Wolniansky,
  ``Simplified processing for high spectral efficiency wireless communication
  employing multi-element arrays,'' \emph{IEEE Journal on Selected areas in
  communications}, vol.~17, no.~11, pp. 1841--1852, 1999.

\end{thebibliography}

\end{document}